\documentclass[conference]{IEEEtran}
\usepackage{xcolor,colortbl}
\usepackage{amsmath}
\usepackage{amssymb}
\usepackage{adjustbox}
\usepackage{color}
\PassOptionsToPackage{hyphens}{url}
\usepackage{url}
\usepackage{graphicx}
\usepackage{epsfig}
\usepackage{latexsym}
\usepackage{enumerate}
\usepackage{balance}
\usepackage{subcaption}
\usepackage{multirow}
\usepackage{soul}
\usepackage{xspace}
\usepackage{booktabs}
\usepackage{wasysym}
\usepackage{xstring}
\usepackage{adjustbox}
\usepackage{mathtools}
\usepackage{physics}
\usepackage{algorithm}
\usepackage{algpseudocode}
\usepackage{algorithmicx}
\usepackage[normalem]{ulem}
\usepackage[T1]{fontenc}
\usepackage{endnotes}
\usepackage{pifont}
\usepackage{comment}
\usepackage{balance}
\usepackage{longtable}
\usepackage{xtab,booktabs}
\usepackage[thinc]{esdiff}
\usepackage{empheq}
\usepackage{xcolor}
\usepackage{footnote}
\makesavenoteenv{tabular}
\makesavenoteenv{table}
\usepackage{tikz}
\usepackage[backend=bibtex,style=numeric-comp,mincrossrefs=99,sorting=none]{biblatex}
\bibliography{xampl}
\usepackage{caption}
\newcommand{\secref}[1]{\mbox{Section~\ref{#1}}}
\newcommand{\ssecref}[1]{\mbox{Section~\ref{#1}}}

\newcommand{\fref}[1]{\mbox{Fig~\ref{#1}}}
\newcommand{\tabref}[1]{\mbox{Table~\ref{#1}}}
\newcommand{\equref}[1]{\mbox{Equation~\ref{#1}}}
\newcommand{\algoref}[1]{\mbox{Algorithm~\ref{#1}}}

\newcommand{\PP}[1]{
\noindent{\bf \IfEndWith{#1}{.}{#1}{#1.}}
}
\newcommand{\PPP}[1]{
\noindent{\bf \IfEndWith{#1}{}{#1}{#1}}
}

\newcommand{\PPPP}[1]{
\indent{\bf \IfEndWith{#1}{}{#1}{#1}}
}
\newcommand{\PPPPP}[1]{
\indent{\bf \IfEndWith{#1}{.}{#1}{#1.}}
}

\newcommand*\circled[1]{\tikz[baseline=(char.base)]{
            \node[shape=circle,fill,inner sep=1pt] (char) {\textcolor{white}{#1}};}}

\newcommand{\cmark}{\ding{51}}%

\algnewcommand\algorithmicinput{\textbf{Input:}}
\algnewcommand\Input{\item[\algorithmicinput]}
\algnewcommand\algorithmicoutput{\textbf{Output:}}
\algnewcommand\Output{\item[\algorithmicoutput]}
\algnewcommand\And{\textbf{and}}
\algnewcommand\Or{\textbf{or}}
\algnewcommand\Not{\textbf{not}}
\algtext*{EndWhile}
\algtext*{EndFor}
\algtext*{EndIf}

\newcommand{\tildefix}{\raise.17ex\hbox{$\scriptstyle\mathtt{\sim}$}}
\usepackage{hyperref}
\definecolor{linkcolour}{rgb}{0,0.2,0.6}
\definecolor{xgreen}{rgb}{0.2,0.6,0.0}
\definecolor{xred}{rgb}{0.7,0.1,0.0}
\definecolor{Gray}{gray}{0.92}
\hypersetup{colorlinks,breaklinks,citecolor=xred, urlcolor=linkcolour, linkcolor=xgreen}
\usepackage{setspace}

\algtext*{EndWhile}
\algtext*{EndFor}
\algtext*{EndIf}

\usepackage{hyperref}
\definecolor{linkcolour}{rgb}{0,0.2,0.6}
\definecolor{xgreen}{rgb}{0.2,0.6,0.0}
\definecolor{xred}{rgb}{0.7,0.1,0.0}
\hypersetup{colorlinks,breaklinks,citecolor=xred, urlcolor=linkcolour, linkcolor=xgreen}
\usepackage{setspace}

\newcommand{\sys}{\mbox{\textsc{Scaphy}}\xspace }
\newcommand{\ic}{\mbox{\textsc{$I_C(s)$}}\xspace }

\newcommand{\physics}{\mbox{\textsc{physics}}\xspace}

\usepackage{titlesec}

\titlespacing*{\section}
{0pt}{2ex plus 1ex minus .2ex}{1.3ex plus .2ex}
\titlespacing*{\subsection}
{0pt}{1.5ex plus 1ex minus .2ex}{1.3ex plus .2ex}

\newcommand{\squishlist}{
\vspace{5pt}
\begin{enumerate}[noitemsep,nolistsep,leftmargin=.3in]
\setlength{\itemsep}{2pt}
}
\newcommand{\squishend}{
\end{enumerate}
\vspace{5pt}
}

\newcommand{\squishitems}{
\vspace{2pt}
\begin{itemize}[noitemsep,nolistsep,leftmargin=.3in]
\setlength{\itemsep}{0.5pt}
}
\newcommand{\squishitemsend}{
\end{itemize}
}

\newcommand{\attacks}{\mbox{\textsc{40}}\xspace }
\newcommand{\scenarios}{\mbox{\textsc{24}}\xspace }
\newcommand{\industries}{\mbox{\textsc{4}}\xspace }

\newcommand{\accuracy}{\mbox{\textsc{95\%}}\xspace }
\newcommand{\fp}{\mbox{\textsc{3.5\%}}\xspace }
\newcommand{\existfp}{\mbox{\textsc{25\%}}\xspace }
\newcommand{\existaccuracy}{\mbox{\textsc{47.5\%}}\xspace }
\newcommand{\secondexistaccuracy}{\mbox{\textsc{27.5\%}}\xspace }
\newcommand{\secondexistfp}{\mbox{\textsc{12.3\%}}\xspace }

\newcommand{\alter}{\mbox{\emph{process-altering}}\xspace }
\newcommand{\monitor}{\mbox{\emph{process-monitoring}}\xspace }
\newcommand{\init}{\mbox{\emph{initialization}}\xspace }
\newcommand{\control}{\mbox{\emph{process-control}}\xspace }

\newcommand{\ignore}[1]{}

\newcolumntype{L}[1]{>{\raggedright\let\newline\\\arraybackslash\hspace{0pt}}m{#1}}
\newcolumntype{C}[1]{>{\centering\let\newline\\\arraybackslash\hspace{0pt}}m{#1}}
\newcolumntype{R}[1]{>{\raggedleft\let\newline\\\arraybackslash\hspace{0pt}}m{#1}}
\begin{document}

\newcommand{\mr}[1]{\noindent{#1}}

\title{\sys: Detecting Modern ICS Attacks by Correlating Behaviors in SCADA and PHYsical}
\author{
\IEEEauthorblockN{
Moses Ike\IEEEauthorrefmark{2}\IEEEauthorrefmark{3},
Kandy Phan\IEEEauthorrefmark{3},
Keaton Sadoski\IEEEauthorrefmark{3},
Romuald Valme\IEEEauthorrefmark{3},
Wenke Lee\IEEEauthorrefmark{2}
}
\IEEEauthorrefmark{2}Georgia Institute of Technology,
\ \IEEEauthorrefmark{3}Sandia National Laboratories \\
 Email: \{mike, kphan, dksados, rvalme\}@sandia.gov, mosesjike@gatech.edu, wenke@cc.gatech.edu}

\maketitle
\begin{abstract}
Modern Industrial Control Systems (ICS) attacks evade existing tools by using knowledge of ICS processes to blend their activities with benign Supervisory Control and Data Acquisition (SCADA) operation, causing physical world damages. We present \sys to detect ICS attacks in SCADA by leveraging the unique \emph{execution phases} of SCADA to identify the limited set of legitimate behaviors to control the physical world \mr{in different phases}, which differentiates from attacker's activities. For example, it is typical for SCADA to setup ICS device objects during \emph{initialization}, but anomalous during \emph{process-control}.
To extract unique behaviors of SCADA execution phases,
\sys first leverages open ICS conventions to generate a novel physical process dependency and impact graph (PDIG) to identify disruptive physical states. \sys then uses PDIG to inform a \emph{physical process-aware} dynamic analysis, whereby code paths of SCADA \emph{process-control} execution is induced to reveal API call behaviors unique to legitimate \emph{process-control} phases. \mr{Using this established behavior}, \sys selectively monitors attacker's physical world-targeted activities that violates legitimate \emph{process-control} behaviors. We evaluated \sys at a U.S. national lab ICS testbed environment.
Using diverse ICS deployment scenarios and attacks across \industries ICS industries, \sys achieved \accuracy accuracy \& \fp false positives (FP), compared to \existaccuracy accuracy and \existfp FP of existing work.
We analyze \sys's resilience to futuristic attacks where attacker knows our approach.

\end{abstract}
\IEEEpeerreviewmaketitle
\section{Introduction}
\label{sec:introduction}
Unlike Information Technology (IT) attacks, Industrial Control System (ICS) attacks cause physical damages to life-dependent physical processes such as power and water supply. ICS processes are controlled by Supervisory Control and Data Acquisition (SCADA) hosts, which run special programs to control the physical world~\cite{industroyer2, sok, telemetry}. Modern ICS attacks~\cite{florida,industroyer, stuxnet, colonial} are launched from SCADA, where attackers utilize legitimate ICS resources to blend their activities with benign SCADA operations and send malicious signal to disrupt processes.

To detect ICS attacks, statistical analysis of ICS traffic~\cite{dnp_attack, celine, telemetry, modbusmodel, justtraffic, stats1, ml, topologychanges,ocsvm1} are effective against \emph{noisy} behaviors (e.g., network scans and malformed protocols), but are evaded by modern attacks which use legitimate protocols and knowledge of ICS parameters to cause targeted (not noisy) disruptions~\cite{industroyer, industroyer2, sok}.
In addition, physical models monitor sensors to know when \emph{observed} physical states deviate from \emph{expected}, by fitting historical sensor data into \emph{linear} models~\cite{state, limiting, ar}. However,~\cite{ssa,debunk} show that in practice, such models may require experts to build, and a detailed process model may be unavailable. Further, physical models trigger false alarms when deployed in production due to noise and configuration changes, such that benign states appear outside the model~\cite{ssa, limiting}.
In general, existing ICS tools are evaded by modern attacks and prone to false alarms due to analyzing traffic/sensor data in isolation, and therefore they cannot tie their analysis to attack-execution context in SCADA.
Detecting ICS attacks in SCADA is hard because attackers use the same API call behavior as benign SCADA programs.
For example, Industroyer malware, which shutdown Ukraine power grid~\cite{industroyer, industroyer2}, performed malicious actions that are part of normal SCADA activity such as accessing ICS device objects.
Similarly, 2021 Florida water poisoning attack~\cite{florida} used normal Human Machine Interface (HMI) commands to dump toxic chemicals into the water supply.
Hence, existing host agents that looks for \emph{"non-SCADA"} APIs will not detect them.

We found that while these attack behaviors are normal SCADA activities, they are anomalous when performed in atypical \emph{execution phases} in SCADA. Therefore, in this work, instead of treating SCADA as one monolithic execution, we \emph{specialize} its behaviors in unique execution phases, which are \init and \control.
We observe that for Industroyer attack to work, the attacker had to execute API calls that are \emph{atypical} of process-control but needed for the attack. For example, to hijack ICS device handles, Industroyer executed Registry Setup APIs in process-control, which is typical for \init, hence anomalous.
\mr{After infecting SCADA, attackers must "setup/connect" their tools to attack the physical world. These attack behaviors do not align with SCADA's phase-based behavior and leads to atypical APIs in wrong phases.}
Hence if we identify the limited set of legitimate process-control behaviors, we can selectively monitor and detect attacker's physical-targeted activities that violate them. 

Further, because SCADA responds to physical world changes, we can induce SCADA to reveal process-control behaviors by stimulating relevant changes. However, this requires a physical model of ICS processes and their elements' states. Interestingly, we found that we can leverage ICS open platform communications (OPC) conventions~\cite{opc,opc2,opc3} to characterize processes via ICS element configurations. We can then \emph{toggle} each element state to induce SCADA to exhibit its process-control behaviors, enabling us to identify them. Further, we can derive the \emph{impact} of these state changes to identify \emph{disruptive} process states.
This can allow us to detect disruptive physical world effects caused by (state-changing) control signals.

We present \sys, a new hybrid technique to detect ICS attacks by correlating SCADA execution phase-specific behaviors with physical world impact.
\sys identifies the limited set of API calls unique to each SCADA execution phase, which differentiates from attacker's activities in these phases, allowing \sys to detect them.
To characterize the "required" internal host \emph{channels} SCADA uses to control the physical world, we introduce a new reference model, \emph{SCADA Software Stack} ($S^3$), which \sys leverages to selectively monitor steps along attacker's physical world-bound activities \mr{in each execution phases}.
Through $S^3$, \sys can detect attacks that \emph{circumvent} proper $S^3$ layers (e.g., SCADA rootkits) but sends disruptive control signals to the physical world.

\sys uses a physical model to identify disruptive control signals sent to the physical world. \sys generates this model by leveraging OPC conventions (\emph{tags} and \emph{alarms}) to extract and map ICS elements to their processes based on a novel process dependency and impact graph (PDIG) model. PDIG enables \sys to assign each element state an \emph{impact coefficient} (\ic) based on how they impact (decrease or increase) process outcomes. Further, \sys leverages PDIG to help induce and extract legitimate process-control behaviors. To do this, \sys performs a \emph{physical process-aware} dynamic analysis, whereby a SCADA engine~\cite{factoryio} is induced to execute process-control code paths by iteratively switching ICS element states connected to the process. During this, \sys records executed API calls to establish a set of PHYSical world Impact Call Specialization (\physics) constraints to identify \emph{legitimate} process-control behaviors. 

\sys detects modern ICS attacks that are missed by existing tools. By correlating behaviors in SCADA and physical, \sys provides contextual alerts for ICS operators to respond to attacks at both SCADA and physical plant. 
Via \physics constraints, \sys limits the operations an attacker can execute to disrupt a physical process by detecting attacker's anomalous API behavior in atypical execution phases.
\sys's physical model detects when control signals cause a physical process to have \emph{inconsistent state} or driven outside its \emph{setpoint} ranges.
\sys's use of open OPC convention makes it device agnostic and not based on any device/controller, making it work for any OPC-supported ICS deployment.
We evaluated \sys at a U.S. national lab state-of-the-art ICS testbed.
We launched \attacks attacks on \scenarios diverse ICS scenarios across \industries industries, including an open-source Texas Pan Handle power grid~\cite{texas}. \sys detected \accuracy of all attacks with only \fp false positives, including real world ICS malware attacks.
We make the following contributions:
\begin{enumerate}
  \item We propose a hybrid technique to detect ICS attacks by correlating SCADA behaviors with physical world effects.
  \item We present an ICS physical model via OPC conventions to both identify disruptive physical states and extract legitimate process-control phase behaviors in SCADA.
  \item We introduce a new reference model in ICS, \emph{SCADA Software Stack} ($S^3$), to characterize internal software \& hardware layers of SCADA operation. Through $S^{3}$, host agents can selectively monitor $S^3$ layers to detect attacks.
  \item Using diverse ICS scenarios \& attacks, \sys achieved \accuracy accuracy and \fp false positives (FP), compared to \existaccuracy accuracy and \existfp FP of existing work~\cite{state, telemetry}.
  \item Due to limited resources and datasets for diverse ICS security research~\cite{swat2,swat1, morris1, morris2}, we make available diverse ICS experiment scenarios\footnote{https://github.com/lordmoses/SCAPHY} and datasets derived from them, both at the physical sensor layer and from the SCADA components. These experiments can be run in the FactoryIO ICS Engine~\cite{factoryio} and Siemens Step7-based development suite, WinSPS~\cite{winsps}.
\end{enumerate}

\section{\xspace \xspace \xspace Background and Motivation}
\label{sec:overview}
We present a background on ICS and use recent attacks as running examples to motivate our problem; 2021 Florida (FL) water poisoning attack~\cite{florida}, and 2016 Industroyer malware power grid attack~\cite{industroyer2}. In \secref{sec:approach}, we detail \sys's approach in real world settings using the Industroyer example.

\begin{figure}[t]
\centering
\includegraphics[width=0.48\textwidth]{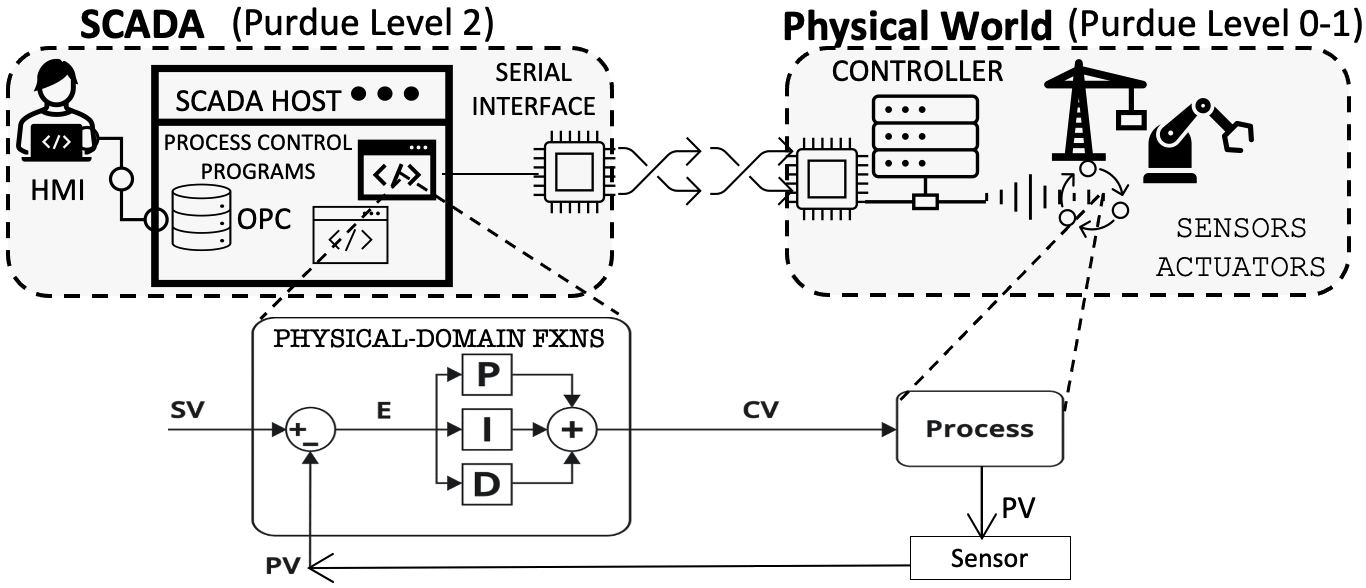}
  \caption{Showing SCADA process-control operation: A physical-domain function compute a control variable \emph{CV} to effect change on the physical world}
\label{fig:layout}
\vspace{-1em}
\end{figure}

\begin{figure*}[t]
\centering
\includegraphics[width=0.75\textwidth]{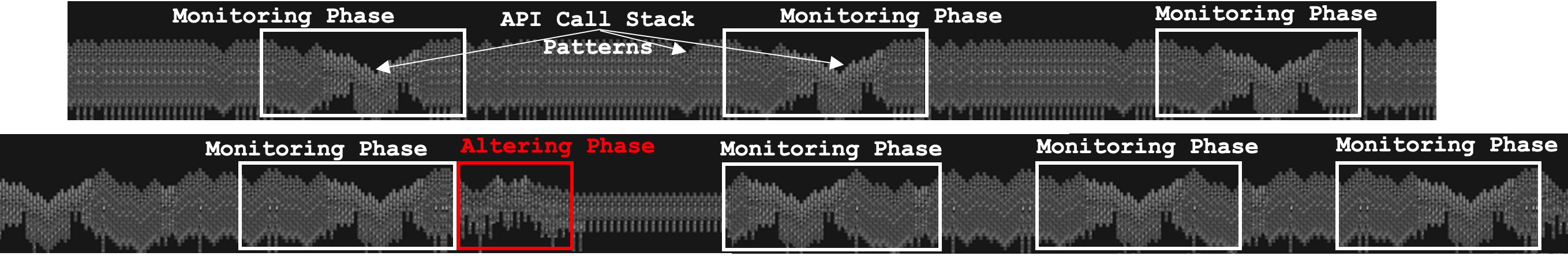}
  \caption{Cycles of Process-Monitoring and Process-Altering phases based on the Step7-based WinSPS~\cite{winsps} control of Water Treatment Plant operation}
\label{fig:wave}
\vspace{-2em}
\end{figure*}
\subsection{Real-World Motivating Examples}
\PP{Florida Water Poisoning Attack}
An attacker raised \emph{dosing rate} of Sodium Hydroxide (NaOH) in FL water treatment plant to toxic levels, endangering citizens. NaOH is used to balance water PH but is toxic in high amounts. After gaining access to SCADA, the attacker started an HMI program to issue attack signals to disrupt the \emph{level control} and \emph{dosing} processes, increasing NaOH from normal 100 ppm to 11,100ppm~\cite{florida, florida2}.\\
\PP{Industroyer Power Grid Attack}
Industroyer shutdown a Ukranian power station by sending malicious signals from a SCADA host to a Siemens SPIROTEC device that runs circuit breaker Remote Terminal Units (RTUs). To perform the attack, Industroyer hijacked host serial COM ports, stole the breaker's \emph{tag} from OPC to \emph{address} its payload, and opened the breakers. This caused power imbalance that shutdown the station~\cite{industroyer, industroyer2}. 

\subsection{ICS/SCADA Operations Background}
\label{ssec:physical-model}
\fref{fig:layout} describes ICS operation based on Purdue model~\cite{purdue}. SCADA hosts at Purdue Level 2 control physical processes, which run at Level 0 and 1 \mr{via ICS elements (comprising of actuators, sensors, and parameters in programmable logic controllers or PLC)}. SCADA constantly monitor running processes and when change is needed, they execute physical-domain logic such as \emph{proportional integral derivative (PID)} to compute a control variable (\emph{CV}) to modify element's states and effect the change (\fref{fig:layout}). For example, the \emph{level control} process in the FL water plant controls chemical level in a tank via a \emph{PID} logic that control how much fluid enters and leaves the tank~\cite{morris2}. This is known as \emph{process-control} and is the main function of SCADA. Unfortunately, ICS attacks are launched from SCADA due to direct access to the physical world as in Industroyer and FL attacks. Attacker gain access to SCADA via IT means such as phishing. Then he can leverage ICS resources such as OPC, HMIs, and SCADA channels to "address" and send malicious signals to ICS elements.\\
\mr{\PPPP{OPC} is a functional part of SCADA and used for interoperability to exchange data about plant information in standardized convention~\cite{opc,opc2,opc3}. OPC is used in widely deployed ICS platforms such
 as Siemens Tia Portal \& Scheider OASYS. For example, Industroyer malware accessed the Ukrainian power grid’s OPC server to extract the circuit breaker’s OPC tag to address its attack payload. In our example, this OPC tag is \emph{BRK.0.BOOL}, which specifies element \emph{BRK} id=\emph{0} with BOOL parameter --- a heuristic about its possible states: $0$ or $1$~\cite{significance}}.

\PP{Limitations of Existing Work} Existing work to detect ICS attacks focus on analyzing traffic and/or sensor data in isolation, without SCADA execution context, which limits their ability to tie their analysis to SCADA for better accuracy and less false alarms.
PLC techniques~\cite{weaselboard,formby} to detect malicious ladder logic on PLCs do not analyze SCADA control signals as in ladder logic but forwards signals to device. \mr{Lee~\cite{lee} only monitors for host DLL injection, which may not happen in ICS attacks.}
\tabref{tbl:taxonomy} is a taxonomy of leading recent works in ICS attack detection, categorized by their technique and data point. We highlight their limitations such as not having ICS diversity in their evaluation and testing in-the-wild (adapted) attacks. We found that this is due to limited resources to emulate end-to-end and diverse ICS security research~\cite{swat2,swat1, morris1, morris2}. \\
\PPPP{ICS Traffic} approaches~\cite{dnp_attack, justtraffic, modbusmodel, functioncode} analyze abnormal traffic flows and protocols. 
However, modern attacks such as Industroyer and FL water plant attacks uses legitimate ICS protocols to emit traffic flows that blends with benign traffic. Traffic timing analysis~\cite{celine,timedetection,telemetry} are effective for analyzing round trip time delays and inter-arrival times but are only effective against attack behaviors that are \emph{chatty}~\cite{celine} such as network scans, but not modern attacks which are targeted.\\
\PPPP{Physical Models} detects deviation from expected physical behavior by training sensor data using linear and auto regressive models~\cite{limiting,state, ar}.
However,~\cite{ssa,debunk} show that in practice, such models may require experts to build, and a detailed process model may be unavailable. Further they trigger false alarms in production due to noise and configuration changes such that benign states are outside the model. 
\mr{Reinforcement and Deep Learning~\cite{rl1,rl2,rl3,rl4} which uses game-theory to learn normal and attack behaviors, requires a high-interaction environment, known attacks, and expert reward function, which may limit its use in diverse ICS practice.} \mr{Process-aware approaches~\cite{bro, runtime-powerflow,process-semantics,process-awareness,runtime-monitoring} focus on specific physical functions, which reduces ambiguity in detection, but requires experts to specify process-specific violations (e.g.,~\cite{bro, process-semantics} uses BRO rules for safety thresholds)}.

\begin{figure*}[t]
\centering
\includegraphics[width=0.85\textwidth]{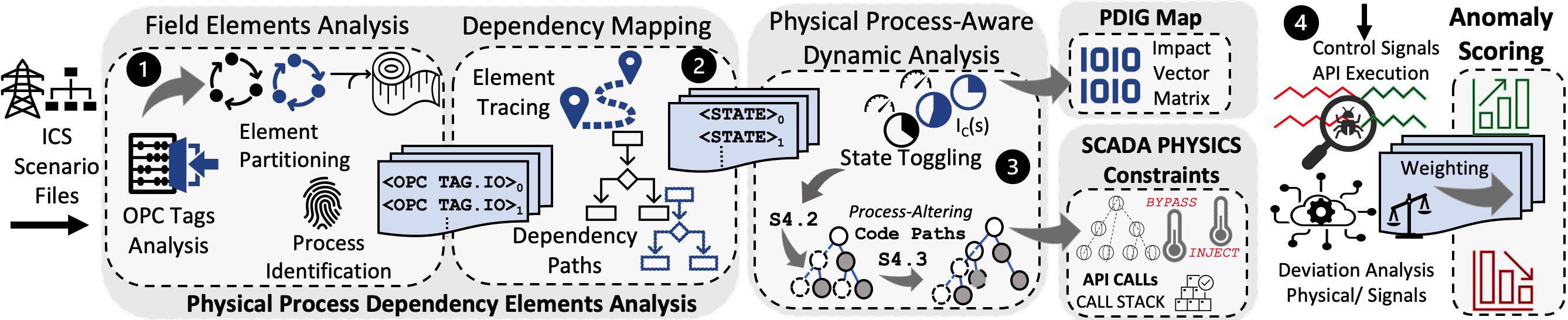}
  \caption{\sys Architecture: ICS scenario file is parsed. Extracted ICS elements are analyzed to identify processes and dependency elements. Element states are toggled to derive impact on process output over scan cycles, during which SCADA process-control \physics constraints are identified for each process.}
\label{fig:system}
\vspace{-2em}
\end{figure*}
\subsection{SCADA Host Attacks and Security Challenges}
Due to physical ramifications of SCADA executions, existing host analysis techniques cannot be directly applied to secure SCADA hosts~\cite{significance, industroyer}.
Due to complex safety constraints in physical tasks, SCADA runs many proprietary and physical domain-specific programs such as ICS drivers and hardware tools. As such, unlike IT programs, using existing concolic analysis tools~\cite{forecast, s2e, qsym} will be intractable due to hardware-constrained code paths and environment need~\cite{forecast,qsym}. Due to legacy and third-party components, code-signing cannot be strictly enforced to whitelist programs, enabling attackers to modify benign programs. For example, Industroyer executed custom APIs, and Stuxnet~\cite{stuxnet} injected into Siemens programs.\\
\PPPPP{Our Insight} We found that the nature of physical tasks, which occur in cycles of repeated steps, requires SCADA to exhibit two distinct execution phases: \emph{initialization} and \emph{process-control}. Process-control comprise of \monitor and \alter sub-phases. \mr{After infecting SCADA, attackers must “setup” and “connect” their tools to attack the physical world. These attack activities do not align with SCADA phase-specific operation, hence results in API calls executed in inappropriate SCADA phases. For example, to access target device tags, Industroyer made OPC calls in the process-monitoring phase, but OPC calls are used in \init. To hijack SCADA physical channels (from benign program), Industroyer and Havex malware~\cite{havex} created ICS device objects while in the process-altering phase, but that was a process-monitoring behavior, hence indicative of an attack. Therefore, \sys deems API calls as “anomalous” when executed in atypical SCADA execution phase}.
\noindent\setlength\tabcolsep{3pt}\begin{table}[t]
    \centering
    \footnotesize
    \resizebox{0.98\columnwidth}{!}{
        \begin{tabular}{@{}l|lll|lll|lll |lll| llll|l>{\columncolor[gray]{0.9}}l@{}}
            \toprule
            &\multicolumn{3}{c|}{ICS}&\multicolumn{3}{c|}{PLC-}&\multicolumn{10}{c|}{Physical Models}&\multicolumn{2}{c}{Has}\\
            \cline{8-17}
            &\multicolumn{3}{c|}{Traffic} & \multicolumn{3}{c|}{Logic-} &\multicolumn{3}{c|}{Physical} &  \multicolumn{3}{c|}{\mr{Reinforc-}} & \multicolumn{4}{c|}{\mr{Process-}} &\multicolumn{2}{c}{SCADA}    
            \\
            &&& & && &\multicolumn{3}{c|}{State} &  \multicolumn{3}{c|}{\mr{Learning}} & \multicolumn{4}{c|}{\mr{Aware}} &\multicolumn{2}{c}{Context}
            \\
            Technique
            &\rotatebox{90}{Yang~\cite{deeplearning}} &\rotatebox{90}{Ihab~\cite{bayes}}&\rotatebox{90}{Ponoma.~\cite{telemetry}} 
            &\rotatebox{90}{Niang~\cite{formal}} &\rotatebox{90}{Mulder~\cite{weaselboard}}&\rotatebox{90}{Formby~\cite{formby}}
            &\rotatebox{90}{Ghaeini~\cite{state}}  &\rotatebox{90}{Dina~\cite{ar}} &\rotatebox{90}{Aoudi~\cite{ssa}} 
             &\rotatebox{90}{\mr{Kurt}~\cite{rl1}} &\rotatebox{90}{\mr{Zhong}~\cite{rl2}} &\rotatebox{90}{\mr{Panfili}~\cite{rl3}}
            
            &\rotatebox{90}{\mr{Chromik}~\cite{bro}}
            &\rotatebox{90}{\mr{Nivethan}~\cite{process-semantics}}
            &\rotatebox{90}{\mr{Remke}~\cite{process-awareness}}
            &\rotatebox{90}{\mr{Lin}~\cite{runtime-powerflow}}

            &\rotatebox{90}{Lee~\cite{lee}} &\rotatebox{90}{\sys}\\
            \cline{1-19} 
            \textbf{Traffic Analysis} &&&&&&&&&&&&&&&&&&\\
            Protocol fields &$\bullet$ &&$\bullet$&&&&&&&&&&&&&&&$\bullet$ \\
            Time analysis &$\bullet$ &$\bullet$&$\bullet$&&&&&&&&&&&&&&& \\
            \cline{1-19} 
            \textbf{PLC Logic} &&&&&&&&&&&&&&&&&&\\
            Logic execution &&&&&&&&$\bullet$&$\bullet$&&&&&&&&&  \\
            Logic verification &&&&&&&$\bullet$&&&&&&&&&&&\\
            \cline{1-19}  
            \textbf{Physical State} &&&&&&&&&&&&&&&&&&\\
            State Deviation &&& &$\bullet$ &$\bullet$&$\bullet$&&&&&&&&&&&&  \\
            Physical Impact&&&&&&&&&&&&&&&&&&$\bullet$ \\
            \cline{1-19} 
            \textbf{\mr{RL Models}} &&&&&&&&&&&&&&&&&&\\
            \mr{Online POMDP} &&&&&&&&& &\mr{$\bullet$}&& &&&& &&\\
            \mr{Reward weights} &&&&&&&&& &&\mr{$\bullet$}& &&&& &&\\
            \mr{Multi-agent game} &&&&&&&&& &&\mr{$\bullet$}&\mr{$\bullet$} &&&& &&\\
            \cline{1-19}
            \textbf{\mr{Process-Aware}} &&&&&&&&&&&&&&&&&&\\
            \mr{Predefined rules} &&&&&&&&& &&& &\mr{$\bullet$}&\mr{$\bullet$}&\mr{$\bullet$}& &&\\
            \mr{Power-flow specs} &&&&&&&&& &&& &&&\mr{$\bullet$}&\mr{$\bullet$} &&\\
            \mr{Power prediction} &&&&&&&&& &&& &&&&\mr{$\bullet$} &&\\
            \cline{1-19}
            \textbf{SCADA Context} &&&&&&&&&&&&&&&&&&\\
            DLL inject-based &&&&&&&&&&&&&&&&&$\bullet$& \\
            PHYSICS constr.&&&&&&&&&&&&&&&&&&$\bullet$\\
            \cline{1-19}  
            \rowcolor{Gray}
            \textbf{Evaluation} &&&&&&&&&&&&&&&&&&\\
            \cline{1-19}  
           \rowcolor{Gray}
            In-the-wild attacks &&&&&&&$\bullet$&&&&&&&&&&&$\bullet$ \\
            \rowcolor{Gray}
           ICS diversity &1&1&1&1&1&1&1&1&1&1&2&1&1&1&1&1&1&4\\ 
           \bottomrule
        \end{tabular}
    }
    \caption{Taxonomy of Recent Related works in ICS Attack detection}
    \label{tbl:taxonomy}
    \vspace{-1em}
\end{table}

\PP{SCADA Execution Phases} SCADA starts with \init which is performed once to setup environment such as loading ICS drivers. After \init, process-monitoring starts. It involves updating process states in memory and operator HMIs. When physical change is needed, process-monitoring transitions to process-altering to perform the change and return until change is needed again. Process-altering invokes physical-domain logic on a \emph{setpoint} variable \emph{SV} to output a \emph{CV} which is sent to the process (\fref{fig:layout}). \fref{fig:wave} shows the unique execution patterns of process monitoring and process-altering phases of a water treatment operation based on Siemens S7 WinSPS SCADA platform~\cite{winsps}. \mr{This pattern is based on API call stack behavior, which identifies these SCADA phases. }

To identify behaviors unique to each process-control phase, \sys first leverages domain-knowledge in OPC to build a physical model of ICS processes, which enables it to detect physical anomalies of process state changes such as when a process has \emph{inconsistent state}. \sys then uses the model to inform a \emph{physical process-aware} dynamic analysis to induce and extract the limited set of legitimate API calls unique to each process-control phase behaviors. Through the API calls, \sys establishes a PHYSical world Impact Call Specialization (\physics) constraints, which attackers must violate (i.e, execute \emph{atypical} APIs) to attack the physical world.
Further, \sys develops a new reference model, \emph{SCADA Software Stack} ($S^3$) to characterize the "required" internal SCADA \emph{channels} to access the physical based on the execution phases. For example, calling \emph{ReadFile} indicates \monitor and \emph{WriteFile} indicates \alter. Both APIs access ICS device objects (e.g, COM Ports) in 3rd layer of $S^3$, to send signals to physical devices. Through $S^3$, \sys can detect attacks that \emph{circumvent or bypasses} required $S^3$ layers (e.g., SCADA rootkits) but sends disruptive signals to devices. 

\mr{\PP{Resilience Against Evasion in SCADA} To attack the physical world, an attacker must communicate over SCADA's physical interface, through (only) which the physical can be accessed. \sys monitors all network communication in the SCADA host via a hypervisor, which runs at a higher privilege than the SCADA system. \sys enforces that all access to the physical world go through the expected physical world-bound API calls by monitoring accesses to the $S^3$ layers at runtime from the hypervisor using virtual machine introspection (VMI). If an attacker bypasses these expected APIs and sends signal to the physical world, \sys will see that the emitted signal does not have the expected provenance (i.e., no matching API call occurred), and consequently detect the attacker's influence on the system.}

\section {\xspace \xspace \xspace Threat Model and Assumptions}
\label{sec:threat_model}
We assume a threat model similar to existing work for SCADA-originated ICS attacks~\cite{telemetry, celine, ocsvm1, state}, 
whereby attacker has infected SCADA and can use available tools to send attack signals to devices. We developed \sys for the Windows-based SCADA systems, which has unmatched dominance in ICS. We make the following practical assumptions:
We do not consider attacks that do not originate from SCADA such as side-channel~\cite{sc-emf,sc-power}. 
Majority of \emph{in-the-wild} ICS attacks are SCADA-originated~\cite{attacksok2}. Notional malware that \emph{originates}/runs \emph{only} on PLC such as in~\cite{harvey} are rarely seen in the wild due to attacker's cost of developing reusable malware for non-traditional CPUs~\cite{plclock, sok}. We note that PLC Man-In-The-Middle (MITM) has been addressed by existing work~\cite{mitm1, mitm2, mitm3} and in practice via non-PLC diode gateways~\cite{diode}, and hence is outside the scope of this work. \mr{\sys assumes the hypervisor as Trusted Computing Base (TCB) that cannot be compromised. In the hypervisor, \sys sees all physical-bound API calls on SCADA virtual machine (VM). Further, \sys relies on the Windows Kernel Patch Protection (KPP) to detect when rootkits inject into the ICS Device Stack (2nd layer of $S^3$) to MITM legitimate SCADA programs. KPP ensures that third-party kernel drivers (e.g., rootkits) cannot modify the kernel subsystem, which mediates hardware-access (i.e., rootkits cannot hide hardware-bound API calls). Relying on KPP is practical because it is widely used in Windows.}
\section{ \sys Approach}
\label{sec:approach}
\begin{figure*}[t]
\minipage{0.21\textwidth}
\begin{subfigure}[c]{\textwidth}
    \includegraphics[width=0.99\textwidth]{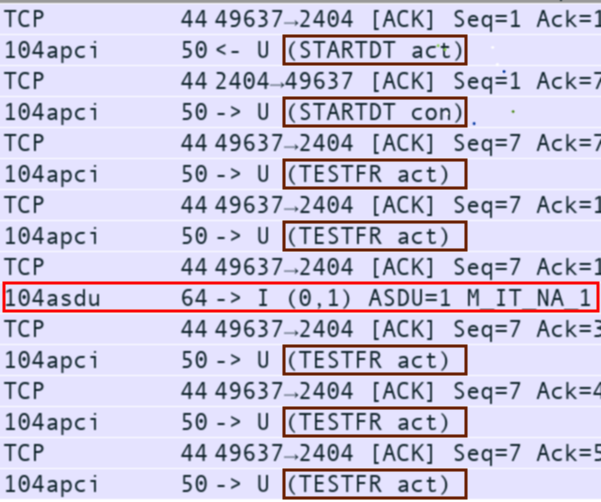}
    \vspace{-3ex}
  \subcaption{Industroyer's IEC Attack Traffic}
    \label{fig:industroyer_traffic}
\end{subfigure}
\endminipage\hfill
\minipage{0.22\textwidth}
\begin{subfigure}[c]{\textwidth}
    \includegraphics[width=0.99\textwidth]{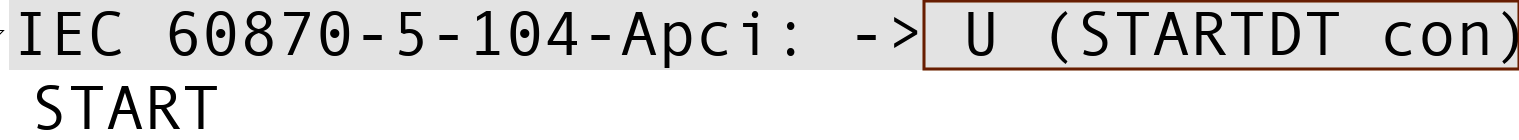}
    \vspace{-3ex}
    \caption{IEC APCI Session START}
    \label{fig:industroyer_start}
 \includegraphics[width=0.99\textwidth]{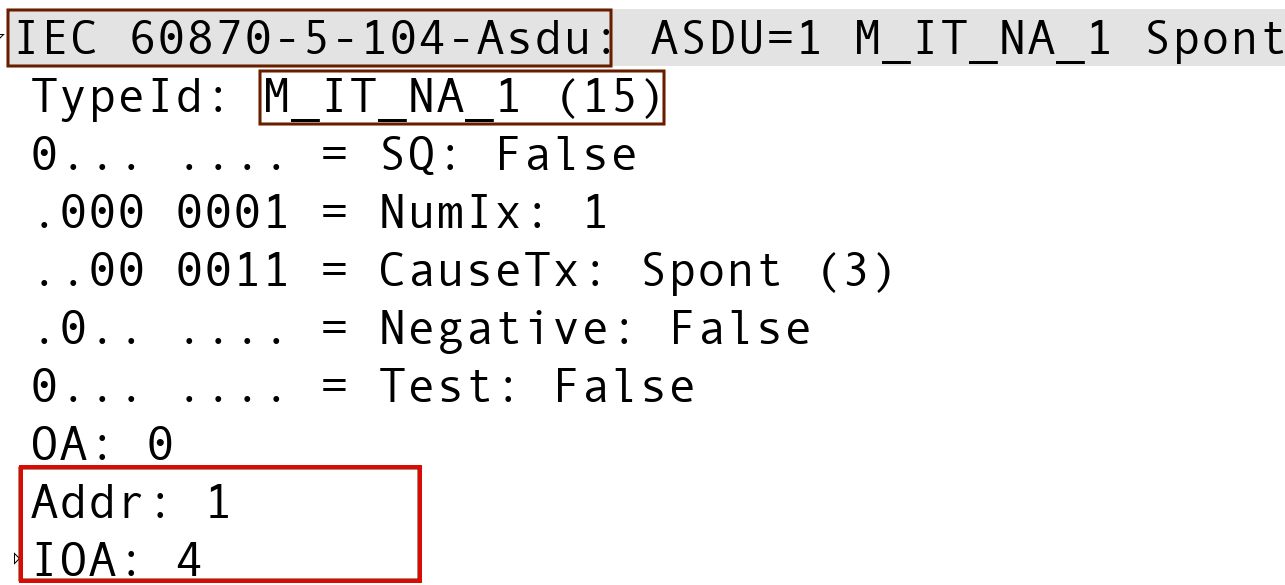}
  \vspace{-3ex}
  \subcaption{Industroyer Payload to Circuit Breaker IOA 4 OPC tag}
    \label{fig:industroyer_payload}
\end{subfigure}
\endminipage\hfill
\minipage{0.26\textwidth}
\begin{subfigure}[c]{\textwidth}
    \includegraphics[width=0.99\textwidth]{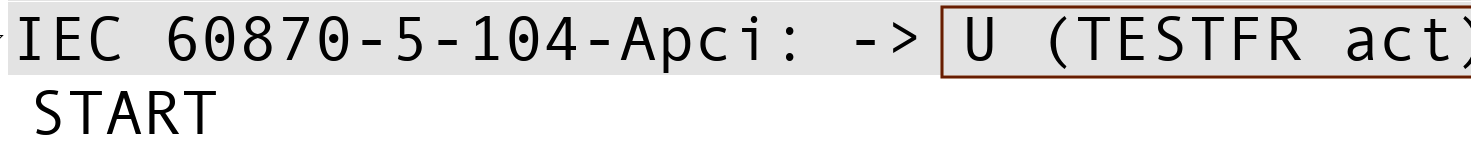}
     \vspace{-3ex}
    \caption{IEC APCI TEST if device is alive}
    \label{fig:industroyer_tester}
    \vspace{-1ex}
 \includegraphics[width=0.99\textwidth]{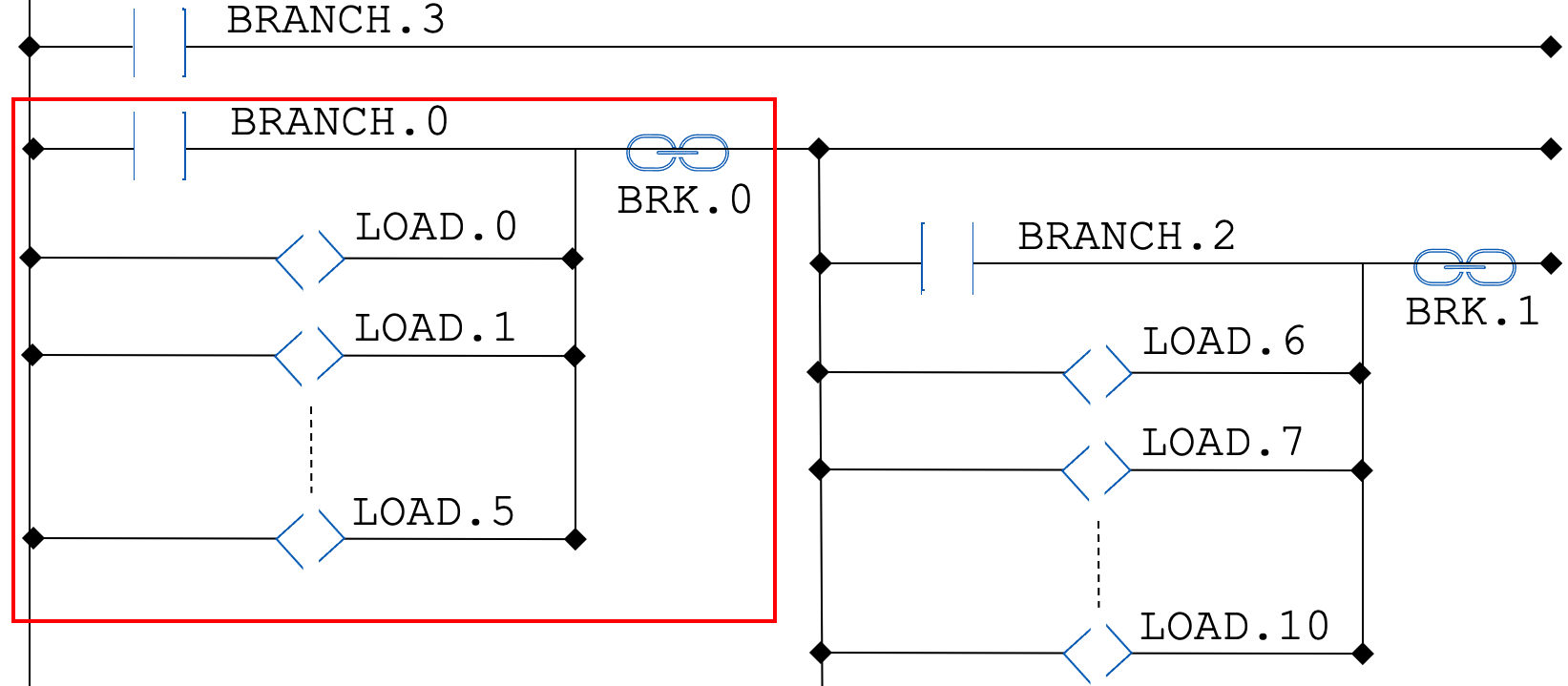}
 	 \vspace{-4ex}
 	 \label{fig:circuit}
    \caption{ Part FBD for Power Load Balancing}
\end{subfigure}
\endminipage\hfill
\minipage{0.29\textwidth}
\begin{subfigure}[c]{\textwidth}
\centering
	\includegraphics[width=0.99\textwidth]{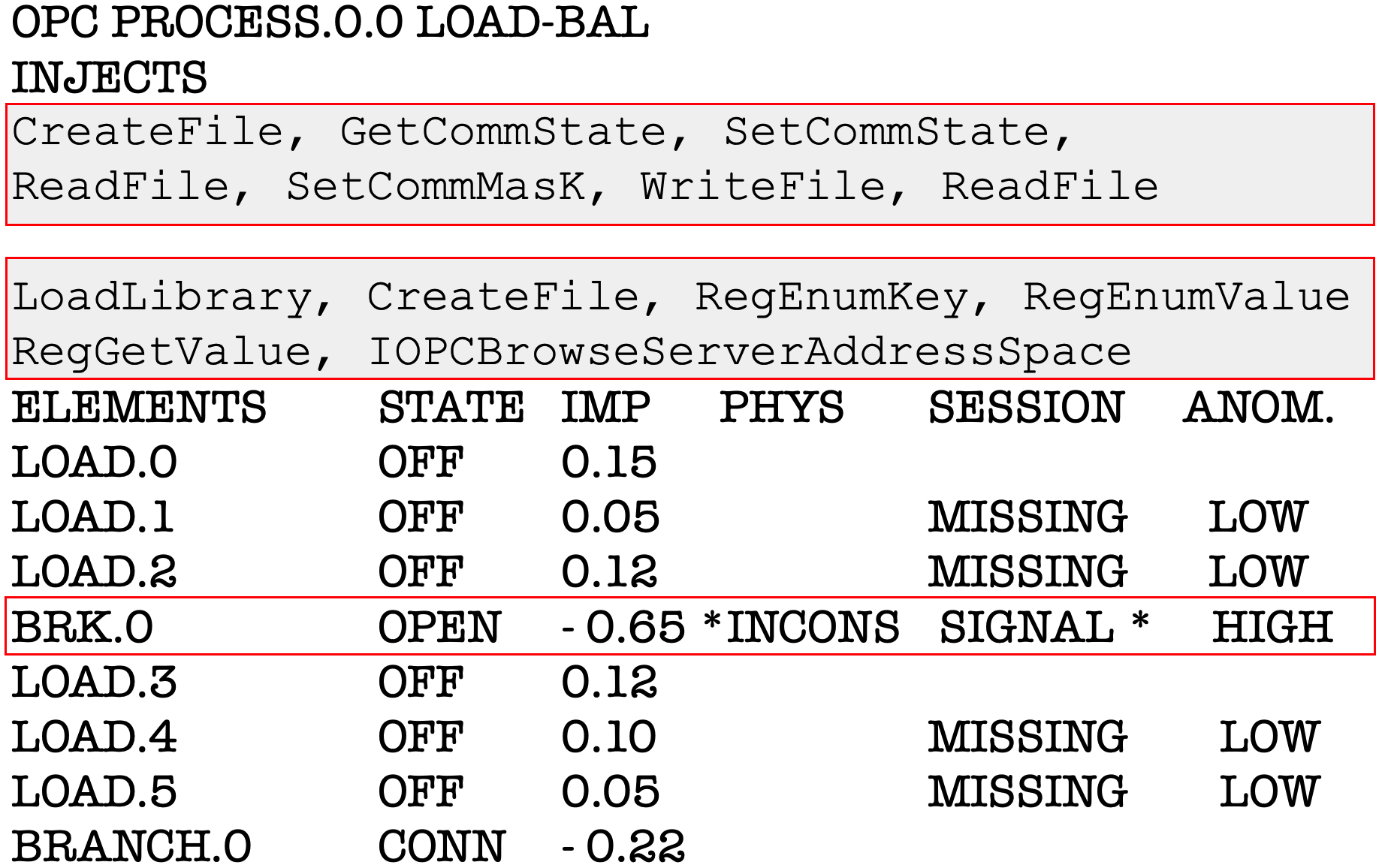}
	\vspace{-3ex}
    \caption{\sys's Industroyer Detection Output}
\label{fig:industroyer_output}
\end{subfigure}
\endminipage\hfill
\vspace{-1em}
\caption{Delineation of Industroyer's power grid attack signals based on the IEC 60870-5-104 ICS Protocol and \sys's SCADA and physical anomaly output}
\vspace{-2em}
  \label{fig:industroyer}
\end{figure*}
\PP{Input and Output}\sys's takes as input, an ICS scenario's OPC element data and function block diagram (FBD) and outputs (i) a physical model and (ii) \physics constraints. The physical model assigns an \emph{impact score} to each element state in a process, based on how the state impacts the process output. \sys uses impact scores to detect signals that are disruptive to a process. \physics constraints are a set of API calls of legitimate \control phase behaviors, which differentiates from attacker's activities, such that he must \emph{inject into} or \emph{circumvent} them to attack processes, allowing \sys to detect it as a violation of \physics constraints.\\
\PPPPP{Deployment} \sys runs in hypervisors of SCADA VMs. Specifically, we deployed \sys in Dom 0 of Xen hypervisor. \mr{This approach is practical because ICS plants are increasingly adopting virtualization to provide redundancy in SCADA such as in Enel power plants~\cite{vm-plant} and recent surveys~\cite{vm1,vm2}. For example, Nozomi Networks, an ICS leader, provides several security solutions (e.g., Vantage and Guardian) that relies on virtualized end points~\cite{vantage, guardian, nozomi-case-studies}.}
In Xen, \sys leverages LibVMI to trace API calls in SCADA VMs. At the network interface, \sys monitor for control signals that changes element states. ICS protocols such as Modbus and DNP3 specifies function code for control signals (e.g., DNP3 $0x02$/Modbus $0x05$). Many traffic analyzers exist to identify function codes in ICS protocols~\cite{vtscada, justtraffic, modbusmodel}.

\subsection{End-to-End Operation} \sys works in four phases as shown in \fref{fig:system}. \circled{1} \sys parses ICS scenario files to analyze and partition ICS elements into terminal and non-terminal sets based on heuristics from OPC. Terminal elements identifies processes.
\circled{2} \sys then traces each element's connections to map dependent element to their process. \sys then loads the scenario in an ICS engine~\cite{factoryio, winsps} and performs a \emph{physical process-aware} dynamic analysis \circled{3}, whereby the engine is induced to execute code paths of \control operations by iteratively switching element states. During this, \sys records the API calls executed during process-altering and process-monitoring phases, to establish PHYSical world Impact Call Specialization (\physics) constraints. Further, the change in process output caused by each state switching is averaged over several scan cycles to derive an impact score for each state relative to others. 
For states with oscillating impact (i.e., process output may increase or decrease), \sys derives a \emph{setpoint range}, which defines a minimum and maximum impact score for the process. \circled{4} \sys raises alarms when executed APIs in each \control phase are not in the \physics constraints, when control signals causes \emph{inconsistent state} or \emph{outside setpoint} anomalies, and when \emph{missing}, \emph{extraneous}, and \emph{out-of-order} signals are seen. \sys then computes an anomaly score.

\subsection{Detecting Industroyer Attack Behavior with \sys}
Industroyer shutdown Ukranian power grid by opening circuit breakers connected to load lines~\cite{industroyer2}. This attack disrupted load balancing of power demand \& supply, a weakness in power systems, leading to outages~\cite{madiot}. To replicate the attack, we adapted an open-source Texas Pan Handle grid~\cite{texas} setup in PowerWorld~\cite{powerworld} at a U.S. National Lab. Our lab setup is detailed in \secref{sec:evaluation}. We ran Industroyer in its "intended" environment; a SCADA host with COM ports and OPC, connected to IEC 608070 device with simulated circuit breaker RTUs. 
\sys raised detection alarm in under 9 seconds for a \physics \emph{injection} violation and an \emph{inconsistent state} anomaly.\\
\PPPP{Industroyer PHYSICS Violation}
Industroyer made several \emph{LoadLibrary} calls, although a normal SCADA API, was performed after the process-altering phase indicated by a prior \emph{CreateFile(Write)} call. We found that \emph{LoadLibray} is normal for process-monitoring and \init but not process-altering, per the \physics constraints established from the power scenario. We found that Industoryer used \emph{LoadLibrary} to load \emph{OPCClientDemo.dll}, to gain OPC capability. Thereafter, Industroyer transitioned to process-monitoring indicated by a \emph{ReadFile} call. It then invoked \emph{IOPCBrowseServerAddressSpace} OPC call to extract circuit breaker Information Object Address (IOA tag) to send its payload as shown in \fref{fig:industroyer_payload}. OPC calls are typical of \init, but not process-monitoring. In addition, Industroyer created new ICS device handles while already in the process-altering phase (to highjack COM Ports) which is malicious in process-altering. \\
\PPPP{Industroyer's Physical Anomalies}
The Industroyer attack generated 8 IEC 608070 signals as shown in \fref{fig:industroyer_traffic}; two IEC 60870 Application Protocol Control Information (ACPI) \emph{START} frames to begin communication, two ACPI \emph{TEST} frames to check controller status, and one Application Service Data Unit (ASDU) payload sent to the circuit breaker RTU. Industroyer also issued 3 last \emph{TEST} signals to verify controller's post attack status as shown in \fref{fig:industroyer_tester}.
Based on the physical model mapping of the element IOA in the payload, \sys identified the target process as \emph{load balancing} (LB). LB's dependency elements are load lines \emph{LOADS.0-5}, breaker \emph{BRK.0}, and \emph{Branch.0} as shown in \fref{fig:circuit}. \sys output (\fref{fig:industroyer_output}) show that Industroyer issued a control signal to \emph{BRK.0} (i.e., indicated by SIGNAL*), with none for other elements (i.e., missing).  However, \emph{BRK.0}'s new open state has an opposing impact vector to load lines's OFF state per \sys physical model, allowing \sys to detect an \emph{inconsistent state} anomaly (detailed in \ssecref{ssec:anomalies}). Using \emph{BRK.0}'s impact vector of 65\%, \sys derived a high anomaly score. 

\section{ System Design}
\label{sec:design}
The goal of \sys is to (\emph{I}) identify the limited set of legitmate API calls of SCADA process-control execution phases that differentiates from attacker activities, and (\emph{II}) build a physical model of ICS processes to identity disruptive physical impact of control signals. To achieve (\emph{I}), we leverage the \mr{process dependency mappings} in (\emph{II}) to inform a \emph{physical process-aware} dynamic analysis. To achieve (\emph{II}), we leverage ICS OPC domain knowledge \mr{and impact scores derived during (\emph{I})} to build a process dependency and impact graph (PDIG).

\subsection{Automated Physical Process Comprehension}
\label{ssec:physical}
\vspace{-.4em}
\PP{OPC Tag Analysis}
\sys parse OPC \emph{tags} of ICS scenarios to discover ICS elements and their parameters. OPC naming conventions allows \sys to extract element's possible states, which allows \sys to automatically \emph{switch} element's states during impact modelling. For example, the extracted OPC tag in Industroyer example \emph{BRK.0.BOOL=1} specifies element \emph{BRK}, ID of 0, and boolean parameter, which tells \sys to switch its state 1$\hookrightarrow$0, as opposed to an integer.

\PP{ICS Element Partitioning and Process Identification}
A process comprises of elements that work together to achieve a physical goal \mr{specified by the plant. This goal is determined by the state of a \emph{terminal} element (a heuristic \sys uses to identify unique processes). For example, the Tank in the Level Control (LC) process of the FL attack example is the terminal element because the LC goal depends on the Tank's level. This level is monitored by a level \emph{Meter} sensor. When Tank's level reaches the plant's specified $SV$ for LC, the process concludes. OPC captures this information via OPC \emph{Alarms\&Event}, a data structure that specifies \emph{monitored} process parameters~\cite{ae,opc-code-event}. To identify terminal elements, \sys analyzes \emph{Alarms\&Event} and extracts process parameter tuples, which corresponds to a sensor and terminal element pair, e.g., the Meter and Tank.}
\sys then partitions ICS elements into two sets; terminal and non-terminal sets ($E_{Term}$, $E_{NTerm})$.
\sys represents a process in the form of $P_j=(S_{j}$, $E_{Term_{j}})$;
where $S_{j}$ is the process sensor that monitors the terminal element $E_{Term_{j}}$, which corresponds to process $P_j$'s output goal. For example, \sys represents LC process as \emph{LC=(Meter,Tank)}. Other elements in the plant such as Pumps and Valves are in $E_{NTerm}$. Based on identifying all processes, \sys partitions the set of elements $E$ such that:
\begingroup
\begin{equation} \label{eq:1}
\small
(E = E_{Term} \cup E_{NTerm})
\end{equation}
\endgroup
\begingroup
\begin{equation} \label{eq:2}
\small
\{\forall i, j \in (E_{Term},E_{NTerm}); i\ne j; i \cap j = \emptyset \}
\end{equation}
\endgroup
\begingroup
\begin{equation} \label{eq:3}
\small
|E_{Term}| := |P|
\end{equation}
\endgroup
$|E_{Term}|$ equals the number of processes, $|P|$

\PP{Process Dependency Mapping}
After identifying processes, \sys traces element's connections to identify dependency elements of a process. \mr{To do this, \sys converts the ICS scenario's function block diagram (FBD) into Statement Lists (STL), a textual representation of FBD logic. STL are network-like statements which connects elements via logic arithmetic. An example STL is shown in (\secref{sec:appendix}). FBD are automatically generated by SCADA programs.} From each $E_{Term}$ node in STL, \sys traces its connection \emph{paths} until all connected elements are identified. Each process (identified by its terminal element) now has a list of paths, \emph{PATHS}, which contain their dependency paths, \emph{DepPath}. \emph{DepPath} is a set of $E_{NTerm}$ nodes arranged in sequential order from the $E_{Term}$ node identifying process $p$, and given as follows:
\begingroup
\begin{equation} \label{eq:4}
\small
DepPath(p) := \{E_{NTerm_{0}},...,E_{NTerm_{n}}\}
\end{equation}
\endgroup
\begingroup
\begin{equation} \label{eq:5}
\small
PATHS(p) := \{DepPath_i(p),...,DepPath_n(p)\}
\end{equation}
\endgroup
\sys then aggregates all elements in all \emph{DepPath} of a process into a dependency element set $DEP(p)$ such that:
\begingroup
\begin{equation} \label{eq:6}
\scalebox{0.95}{%
$\{\forall i \in PATHS(p): DEP(p) := \bigcup_{j=0}^{n}i(p) \}$
}
\end{equation}
\endgroup
$DEP(P)$ is the union of all elements in the dependency paths of $P$'s pathlist. \sys keeps tracks of the internal ordering among these elements as developed from \equref{eq:4}. \sys uses this \emph{ordering} constraints to detect out-of-order signals.\\ 
\PPPP{Functional Inter-Process Relationships} \sys identifies functional relationship between processes with common elements among them. If an element is in $E_{Term}$ of $P_1$ and in $E_{NTerm}$ in $P_2$, then $P_1$ depends on $P_2$. \sys models such element as \emph{inter-process transfer points (PTP)}, and $P_1$ and $P_2$ as \emph{PTP sink} and \emph{PTP source} respectively. PTP instances is common among Boolean elements such as valves and switches. \sys leverages PTP to detect attacks spanning multiple processes, such as in the FL water attack (detailed in \ssecref{ssec:florida}). 
PTP events occur when a control signal causes a PTP element to change state, which causes the PTP \emph{sink} to assume the value of the PTP \emph{source}'s $E_{Term}$. Through this, \sys detects disruptive impact on the PTP \emph{sink} process stemming from the PTP \emph{source}. 

\subsection{Modelling Process Dependency and Impact}
\sys models how each dependency element state of a process impacts (decrease or increase) the process output using a novel \emph{process dependency and impact graph} (PDIG) model. 

\PP{Sketching PDIG Gragh}
PDIG is a set of nodes which represents elements, and edges which connects element nodes based on their relationship. Example PDIG is shown in \secref{sec:appendix} for the LC process. Undirected \emph{process edges} connect $E_{Term}$ elements to each $E_{NTerm}$ element in the \emph{same} $DEP(P)$. Undirected \emph{element edges} connect $E_{NTerm}$ elements without any ordering constraint. Directed \emph{element edges} connect two $E_{NTerm}$ elements with ordering constraints among them in the direction of the ordering. Each $E$ node is annotated with its possible states and impact score (derived later). In the PDIG, \sys identifies and annotates PTP instances when there is an undirected element edge from a terminal element (of the PTP \emph{source}) to a non-terminal element (i.e., the PTP element). 

  \PP{Deriving Impact to Model Physical Effects}
\sys assigns each element state $s$ an impact score \ic based on how they impact a process relative to other states. When a control signal changes an element state, \ic is used to compute the anomaly score to quantify impact. \sys also uses \ic to prune out non-impactful states, which reduces the number of elements to monitor. \mr{ Note that \sys aims to derive “relative” impact score of an element state, not actual score, or score for state combinations. As such, \sys does not consider other elements' state when deriving \ic, which avoids combinatoral state explosion~\cite{criticalstate,pattern1,modbusmodel}. To make the scoring robust/fair, all score derivation starts from the same initial process configuration and is driven until the process reaches steady state.} To derive \ic, \sys leverages an ICS environment; Siemens S7 WinSPS~\cite{winsps} and FactoryIO~\cite{factoryio} to load and drive the ICS scenarios' processes. \sys then iteratively switches each state in the process and analyze the moving average of process outputs via reading the process sensor element. \sys stops evaluating the change in output when successive changes become negligible (less than 1\%). We normalize \ic with respect to scan cycles ran, which bounds its value between 0.0 and 1.0. This succinctly describes the impact of each state relative to other states.\\
\begin{figure}[t]
\minipage{0.40\textwidth}%
\begin{subfigure}[c]{\textwidth}
\centering
  \includegraphics[width=0.99\textwidth]{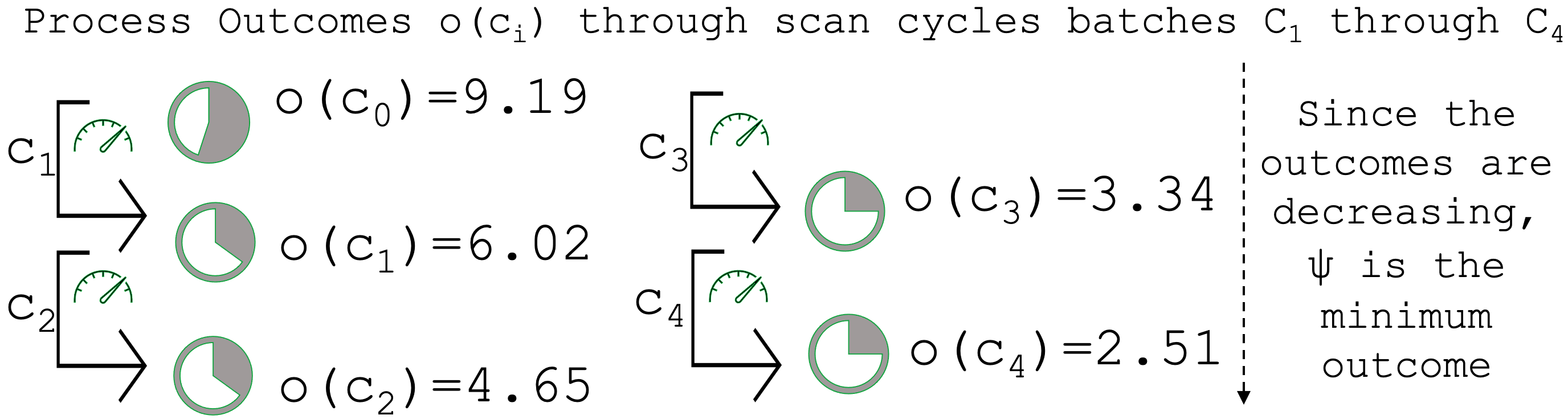}
  \vspace{-3ex}
  \subcaption{Process operation scan cycles}
  \label{fig:ic1}
  \includegraphics[width=0.99\textwidth]{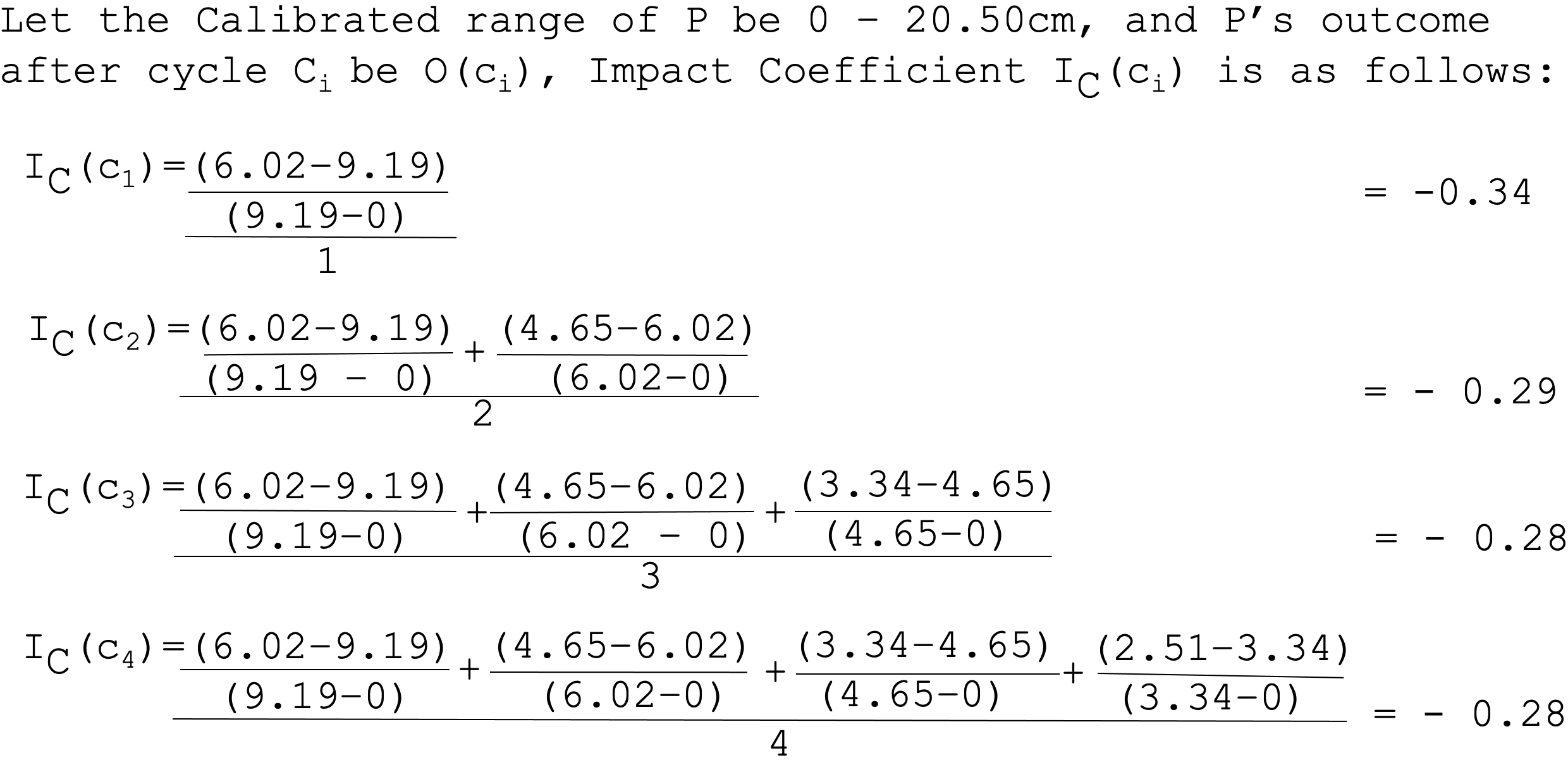}
  \vspace{-3ex}
  \subcaption{Computation of Impact Coefficient}
  \label{fig:ic2}
 \end{subfigure}
 \endminipage
 
 \minipage{0.24\textwidth}
\begin{subfigure}[c]{\textwidth}
  \includegraphics[width=0.99\textwidth]{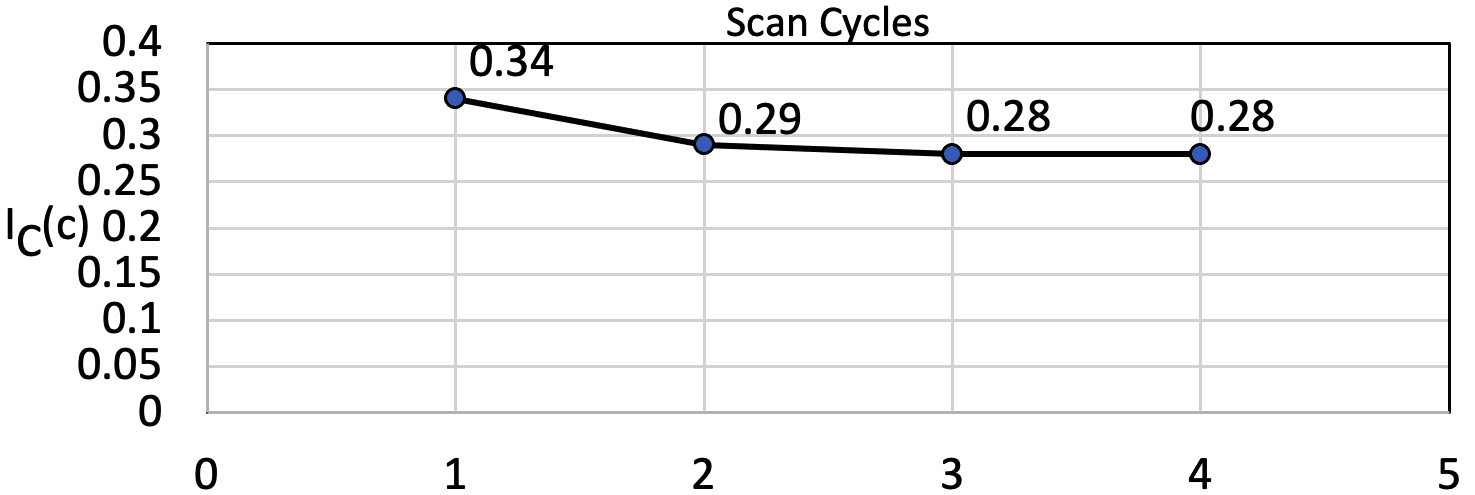}
 \vspace{-1em}
  \caption{A plot of \ic to scan cycles}
  \label{fig:ic3}
  \end{subfigure}
 \endminipage\hfill
  \minipage{0.24\textwidth}
\begin{subfigure}[c]{\textwidth}
 \includegraphics[width=0.99\textwidth]{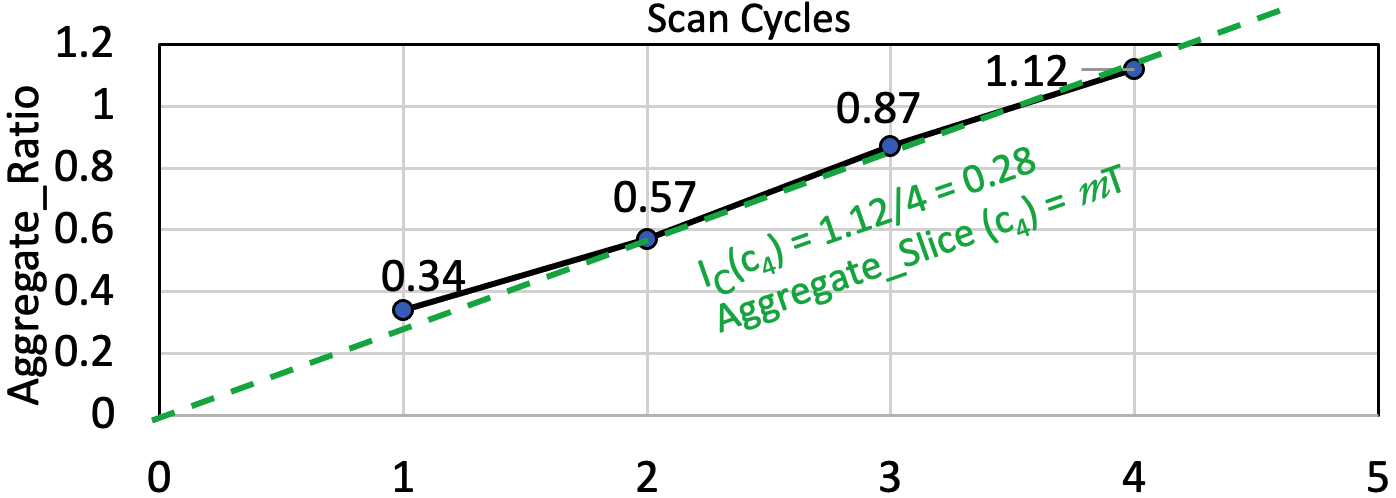}
 \vspace{-1em}
  \caption{Plot of Aggregate Slices to scan cycles}
  \label{fig:ic4}
\end{subfigure}
\endminipage\hfill
 \vspace{-1em}
\caption{\sys Impact Coefficient \ic Derivation for Each Element State}
\vspace{-.5em}
  \label{fig:ic}
\end{figure}
\PPPPP{Formulation of $I_C(s)$}
We define a process \emph{outcome transition} set $\tau_n$, which is a set of ordered outcomes $o$ for a process from scan cycle $c_j$ to $c_n$:
\begingroup
\begin{equation} \label{eq:5}
\small
\tau_n := \{{o(c_j), o(c_{j+1}), o(c_{j+2}), ..., o(c_n)}\}
\end{equation}
\endgroup
where $c_j$ is the first scan cycle following \sys switching of an elemetary state and $c_n$ is the last or current scan cycle observed where $0 \le j \le n, n \in \mathbb{Z}$. \sys keeps track of the highest or lowest recorded outcome $\psi$ (i.e., the boundary outcome). For any scan cycle $c_j$, we define the \ic of the element state under analysis as follows:
\begingroup
\begin{equation}\label{eq:6}
\small
\ic=\frac{\sum\limits_{i =j}^{n} \frac{o(c_{i}) - o(c_{i-1})}{ab(\psi - o(c_{i-1}))}}{|\tau_n|}
\end{equation}
\endgroup
where $o(c_{i}) - o(c_{i-1})$ is change in process outcome, $ab(\psi - o(c_{i})$) is the absolute difference between the highest or lowest outcome and the preceding value at the scan cycle $i-1$. Further, $|\tau_n|$ is cardinality of the scan cycles from $c_j$ to $c_n$.

\begin{figure*}[t]
\centering
\includegraphics[width=0.86\textwidth]{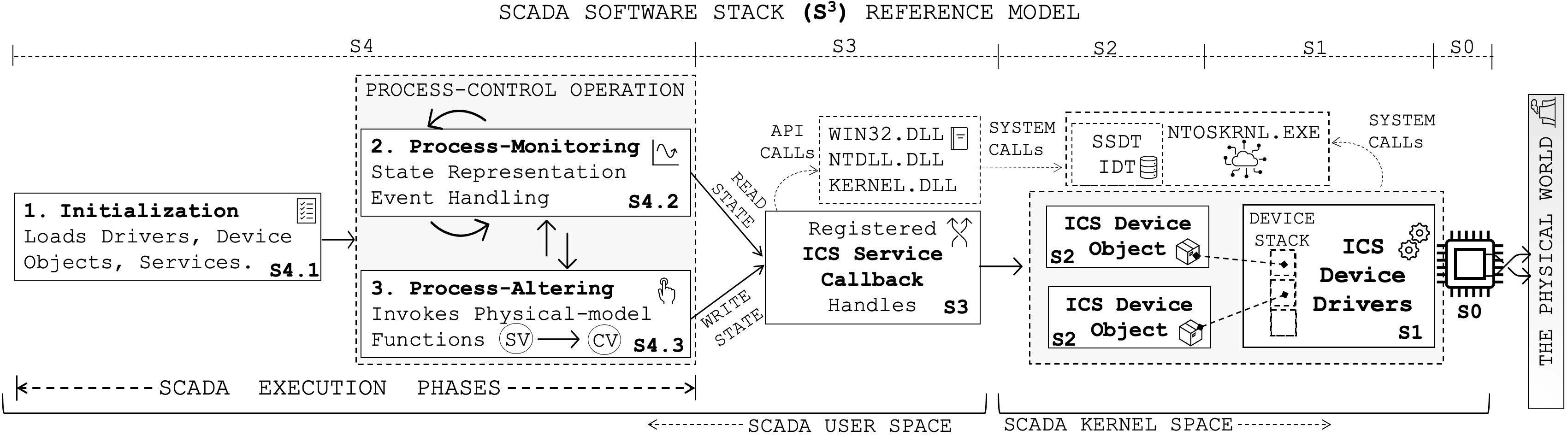}
  \caption{SCADA Software Stack ($S^3$) and SCADA Execution Phases: Showing SCADA Host Interactions with $S^3$ Layers to access the Physical World}
\label{fig:phases}
\vspace{-2em}
\end{figure*}
We see that $\frac{o(c_{i}) - o(c_{i-1})}{ab(\psi - o(c_{i-1}))}$ is the ratio of the current process change (i.e., $o(c_{i}) - o(c_{i-1})$) to maximum change ($\psi - o(c_{i-1})$). If we aggregate this ratio for each scan cycle, we can compute \ic instantaneously at any scan cycles we chose without having to always compute \ic through all previous scan cycles.
Using the aggregate ratio to compute the instantaneous derivation of \ic is given as follows:
\begingroup
\begin{equation} \label{eq:7}
\small
\ic = \fdv{T} Aggregate\_Slice(o(c_n))
\end{equation}
\endgroup
where, for all scan cycles $T$, $Aggregate\_Slice(o(c_n))$ is sum of current change to Max change from $c_j$ to $c_n$, defined as:
\begingroup
\begin{equation} \label{eq:8}
\scalebox{0.99}{%
 $\{\forall c_i \in T: Aggregate\_Slice(o(c_n)):=\sum_{i=j}^{n}\frac{o(c_{i}) - o(c_{i-1})}{\psi - o(c_{i-1})}\}$
 }
\end{equation}
\endgroup
\fref{fig:ic2} illustrates a derivation of \ic through scan cycles $c_1$ to $c_4$. At each scan cycle transition, the generated process outcomes, 9.19 through 2.15 were inputted into the \ic formula to compute the \ic scores. Notice that at scan cycle $c_4$, the difference in the subsequent \ic (i.e., from $c_3$) was negligible (i.e., 0). \sys uses this observation as a heuristic to detect steady states. Otherwise, \sys sets a maximum bound to stop the simulation. \fref{fig:ic4} shows \ic for the end scan cycle $c_n$ ($n=4$) using the iterative form and the instantaneous \ic form (green dotted line).
The $Aggregate\_Slice(o(c_n))$ function is a straight line ($Aggregate\_Slice(c_n)=mT$) drawn from origin to the point $c_n \in T$, where $m$ is the slope. Taking the derivative of $Aggregate\_Slice(c_n)=mT$ gives the instantaneous \ic. \algoref{algo:ic} shows \sys algorithmic approach to derive to \ic. $Aggregate\_Slice$ is the aggregate ratio of measured impact to maximum possible impact, and $Cycles$ is the total cycle batches in terms of scan cycles.

\begin{algorithm}[t]
  \scriptsize
  \caption{ \scriptsize DeriveImpactCoefficient(\ic)}
  \label{algo:ic}
  \begin{algorithmic}
  \Input ElementState $s$, Process $p$
  \Output \ic:
  \smallskip
  \Comment Initialization
  \State $Cycle, Cycle_{MAX} \gets GetCycleBatchAndMax$
  \State $\psi \gets GetOutcomeBoundCalib$
  \State $O_{PREV} \gets GetProcessInitOutcome(p)$
  \State $SteadyState \gets GetSteadyState(p)$
  \State $Aggregate\_Slice \gets 0$
  \While {$Cycle_{MAX} > 0$}
  \State $SDK\_RunSim(s, p)$
  \State $O_{c_i} \gets GetProcessOutcome(p)$
  \State $O_{DIFF} \gets O_{c_i} - O_{PREV}$
  \State $Aggregate\_Slice \gets (Aggregate\_Slice + O_{DIFF})$
  \State $Aggregate\_Slice \gets Aggregate\_Slice/ABS(O_{PREV}) - \psi)$
  \State \ic $\gets Aggregate\_Slice/Cycle$
  \Comment check if steady state is reached, if so return \ic
  \If {$O_{DIFF} < SteadyState$}
  \State Return \ic
  \Else
  \State $Cycle++$
  \State $Cycle_{MAX}--$
  \State $\psi \gets UpdateOutcomeBound(\psi, O_{c_i})$
  \Comment update $\psi$ if neccesary
  \EndIf
  \EndWhile
  \State Return \ic
  \end{algorithmic}
\end{algorithm}

\subsection{Characterizing Physical and Signal Anomalies}
\vspace{-0.5em}
\label{ssec:anomalies}
\PP{Inconsistent State} \sys detects when a process has \emph{inconsistent state} if two dependency element states have \emph{opposing} impact \emph{vectors}. To explain, recall that state's \ic in the PDIG drives process output in \emph{one} direction towards its goal (i.e, not opposing goals). For example, Industroyer attacked \emph{LB} process aims for a \emph{LB} factor of 1.0 between power supply and demand. Although \emph{LB} is affected by other factors, load lines and circuit breaker state play a role. If load decreases/disconnect (e.g., low demand), process-control responds by opening circuit breakers to bring back the balance. Hence, \emph{load disconnecting} is \emph{consistent} with \emph{breaker opening} and drives the process towards its goal. However, \emph{opening} circuit breakers and \emph{connecting} load lines are inconsistent and disruptive to LB and should never occur in any legitimate setting. \sys's \ic model captures this element state relationships and detects such \emph{inconsistent state} physical anomalies. Through this, \sys detected the Industroyer attack based on physical impact of the attack (\secref{sec:approach}).\\
\PPPPP{Outside Setpoint} \sys detects when control signals drive a process to exceed what it is \emph{operationally} caliberated for based on learning the highest and lowest bounds of the process output recovered during the \ic derivation. \\
\PPPPP{Signal Anomalies} \sys detects (i) \emph{missing signals} when a process' control traffic has incomplete signals based on number of its dependency elements. This is useful for targeted attack signals (ii) \emph{Extraneous signals} is when control traffic contains signals for elements not associated with the process. (iii) \emph{Out-of-Order signals} occur signals sequences in control traffic are not in the expected ordered flow based on PDIG.\\ 
\PPPPP{Scoring Anomalies}
\sys computes anomaly scores based on element's \ic. Let $m$ be number of affected elements, and $s_i$ is the state transitioned by the control signal. Let $n$ be number of process dependency elements and $I_{C_{MAX}}(j)$ be the Max \ic for an element. Anomaly score is given by:
\begingroup
\vspace{-.5em}
\begin{equation} \label{eq:8}
\scalebox{.95}{%
$\{\forall j \in DEP(p): Anomaly\_Score = \frac{\sum\limits_{i =0}^{m} I_C(s_i)}{\sum\limits_{j =0}^{n} I_{C_{MAX}(j)}}\}$}
\vspace{-1em}
\end{equation}
\endgroup

Using $I_{C_{MAX}}$ in the denominator normalizes and bounds the scores between 0,1 which succinctly captures the deviation relative to other elements. We then calibrated a detection threshold boundary for low, medium, high anomalies as 0.0-0.25, 0.26-0.60, and 0.61-1.0 respectfully. These gave the best accuracy for all scenarios tested. However, operators can fine tune these detection thresholds as needed. 

\subsection{Analyzing Physical World-Targeted Executions in SCADA}
Given an ICS scenario, \sys aims to generate the limited set of API calls unique to legitimate SCADA process-control operations, referred to as PHYSical world Impact Call Specialization (\physics) constraints. To do this, \sys leverages its physical model to inform a \emph{physical process-aware} dynamic analysis, whereby a SCADA engine is induced to execute code paths of process-monitoring and altering behaviors. However, this requires first knowing each phase identifier and boundary.

\PP{Leveraging SCADA Software Stack ($S^3$) to Characterize Process-Control Behaviors} 
Through in-depth analysis of process-control in diverse ICS settings, we introduce a new reference model, \emph{SCADA Software Stack} ($S^3$). $S^3$ does not replace Purdue Levels~\cite{purdue}. 
Rather, it breakdown Purdue Level 2 (i.e., control systems) into 5 layers to characterize the internal host layers involved in SCADA process-control, shown in \fref{fig:phases}. We hope that via $S^3$, Antivirus companies can develop SCADA-specific host agents to monitor accesses to specific $S^3$ layers to detect attacks. 

SCADA programs (\textbf{$S4$}) do not access ICS devices directly but do so using \emph{device objects} (\textbf{$S2$}), which are software handles that enable the OS to mediate access to physical I/O (\textbf{$S0$}). To support diverse ICS devices, Windows provides a Driver Model (WDM) to allow device vendors to run ICS drivers (\textbf{$S1$}) in the kernel. In WDM, \emph{driver objects} of an ICS driver represent instances of ICS devices the driver supports. For example, Windows supports 16550 UART devices via Windows Serial Driver, which allows SCADA programs to declare device objects, called COM ports, to communicate with devices.

\PP{ICS Callback Functions} To access devices, SCADA programs invoke ICS callback functions (\textbf{$S3$}) registered during ICS driver load. Callback functions invoke \emph{CreateFile} which returns the device object handle as shown in \fref{fig:phases}. Parameters \emph{lpFileName} specifies device object name (e.g., COM1); \emph{dwDesiredAccess} specifies \emph{Read/Write} access mode; \emph{dwShareMode} enables SCADA programs to deny other programs (e.g., malware) access to devices. Unfortunately, attackers (e.g., Havex, Industroyer, Oldrea) subvert this access control by killing SCADA processes to release their handles. For example, Industroyer killed \emph{D2MultiCommService.exe} to hijack all COM ports to Siemens SPIROTEC device. Havex~\cite{havex} scanned COM Ports to discover connected devices. After obtaining device handles, SCADA programs (and attackers) invoke \emph{ReadFile} to read device states or \emph{WriteFile} to send signals to them. Based on this behavior of $S^3$ layers, \sys can monitor \emph{CreateFile}, [\emph{WriteFile} | \emph{ReadFile}] \mr{on device objects} as \emph{identifiers} of process-altering and process-monitoring, respectively.

\PP{Identifying Process-Control Phase Windows}
 Existing work for web servers~\cite{temporal} rely on developer-supplied \emph{boundaries} to identify phase transitions. This manual approach will not work if boundaries are not available such as in proprietary settings like ICS.
 However, \sys leverages $S^3$ layers to analyze the \emph{cyclical} nature of SCADA process-control to identify its transitions from process-monitoring to process-altering. Recall that accessing device objects ($S2$) using \emph{CreateFile} in "Write" mode identifies process-altering, and in "Read" mode identifies process-monitoring, but we need to know when they begin and end (i.e., phase window) to specialize the extracted API calls. Based on their cyclical API call stack behavior such as shown in \fref{fig:wave}, we found that process-altering \emph{follows} process-monitoring. We also found that SCADA performs memory freeing operations thereafter to free up memory buffers used in physical-domain computations and then returns to monitoring.
 
 We analyze process-monitoring to know its phase window. Process-Monitoring comprise of two sub-phases; \emph{process state representation} (reads element states), and \emph{event handling}, which checks if change is needed, otherwise it repeats as shown in \fref{fig:wave}. \sys analyzes the changing \emph{EIP} register and call stack depth of the repeating loop to know the end of event handling the first time it returns to the \emph{EIP} it started. When event handling ends but \emph{EIP} returns to a different location and call stack depth, \sys identifies this as the start of process-altering. Finally, \sys performs a check to confirm the expected $S^3$ identifiers in each phase, which are [\emph{CreateFile}, \emph{ReadFile} || \emph{WriteFile}], \emph{CloseHandle} call sequences for process-monitoring \& process-altering respectfully.
 \begin{table*}[t]
    \centering
    \footnotesize
    \resizebox{0.80\textwidth}{!}{
        \begin{tabular}{@{}l|c|lcl>{\columncolor[gray]{0.9}}c>{\columncolor[gray]{0.9}}l|>{\columncolor[gray]{0.9}}l>{\columncolor[gray]{0.9}}l>{\columncolor[gray]{0.9}}l>{\columncolor[gray]{0.9}}l|l|lccc@{}}
            \toprule
            && \multicolumn{5}{c|}{Physical World Dependency \& Impact Model } & \multicolumn{4}{c|}{PHYSICS Constraints} & & \multicolumn{3}{c}{ICS program files}\\
              ICS&Industry&physical &  &\emph{OPC}&\emph{PDIG} &$I_C$ & \multicolumn{2}{c|}{behavior}& \multicolumn{2}{c|}{verify}   &time & \multicolumn{3}{c}{OPC points and tags}\\
            scenarios &Domain&processes &ID & $E_{Term}$   &nodes  &avg/max  & calls & stack & TP & FP & (min) & size & element & wires \\
            \cline{1-15} 
             
             &  &load balancing&1.1 &c-breaker &6 &.71/.76 &6 &7 &6 &0 &9.3 &9K &19 &35 \\
            \multirow{-2}{*}{power grid} &\multirow{-2}{*}{Power Plant} &pwr distribution&1.2 &load lines &4 &.59/.66 &4 &6 &4 &0 &10.4 &9K &19 &35 \\
            
            \rowcolor{Gray}
            &&level control&2.1 &holding tank &4 &.56/.86 &10 &12 &10 &0 &8.2 &11.5K &13 &23 \\
            \rowcolor{Gray}
           \multirow{-2}{*}{water treatment}&\multirow{-2}{*}{Water Plant} & dosing&2.2 &dose valve &2 &.66/.86 &4 &11 &4 &0 &8.9 &11.5K &4 &23 \\
            
            &&pallet alignment&3.2 &Axes X,Z &6 &.47/.62 &8 &13 &8 &0 &5.9 &9K &10 &11 \\
           \multirow{-2}{*}{auto warehouse}&\multirow{-2}{*}{manufacture} & throughput&3.2 &entry conveyor &2 &.7/.84 &4 &11 &3 &1 &5.5 &9K &6 &11 \\
           
           \rowcolor{Gray}
           &&product quality&4.1 &clamp lid/base &2 &.67/.77 &4 &7 &4 &0 &7.5 &9.5K &8 &19 \\
            \rowcolor{Gray}
           \multirow{-2}{*}{assembler}&\multirow{-2}{*}{manufacture} & load balancing&4.2 &conveyor2 &5 &.8/.84 &6 &14 &6 &0 &7.3 &9.5K &11 &19 \\
            
            &&load alignment&5.1 &push clamp &3 &.47/.71 &7 &13 &7 &0 &5.9 &7.8K &7 &13 \\
           \multirow{-2}{*}{palletizer}&\multirow{-2}{*}{Shipping} & prod protection&5.2 &entry conveyor &6 &.32/.69 &4 &17 &4 &0 &5.7 &7.8K &9 &13 \\
            
           \rowcolor{Gray}
           &&heat setpoint &6.1&room space &3 &.46/.79 &8 &12 &7 &1 &6.1 &6K &8 &14 \\
            \rowcolor{Gray}
           \multirow{-2}{*}{hvac system}&\multirow{-2}{*}{manufacture} & heat flow&6.2 &vent &3 &.6/.82 &6 &19 &6 &0 &6 &6K &7 &14 \\
           
            &&path throughput&7.1 &load/unload &2 &.47/.6 &7 &13 &7 &0 &6.9 &6.2K &9 &15 \\
           \multirow{-2}{*}{converge station} &\multirow{-2}{*}{Shipping} & alt throughput&7.2 &transfer &3 &.67/.88 &6 &12 &5 &1 &6.9 &6.2K &9 &15 \\
           
           \rowcolor{Gray}
           &&alignment&8.1 &control arm &2 &.8/.8 &6 &15 &5 &1 &8.6 &8.5K &11 &17 \\
            \rowcolor{Gray}
           \multirow{-2}{*}{production line}&\multirow{-2}{*}{manufacture} & throughput&8.2 &conveyors &4 &.6/.72 &6 &16 &6 &0 &8.1 &8.5K &11 &17 \\
            
            &&sort accuracy&9.1 &unloader &2 &.54/.85 &4 &13 &4 &0 &5.7 &9K &6 &14 \\
           \multirow{-2}{*}{sort station}&\multirow{-2}{*}{Shipping} & throughput&9.2 &conveyor &7 &.5/.9 &6 &17 &6 &0 &5 &9K &16 &14 \\
           
           \rowcolor{Gray}
           &&accuracy &10.1 &pusher1-2 &6 &.53/.72 &5 &12 &4 &1 &5.9 &4.9K &17 &19 \\
            \rowcolor{Gray}
           \multirow{-2}{*}{separator}&\multirow{-2}{*}{Shipping} & throughput&10.2 &conveyors &7 &.69/.81 &4 &8 &4 &0 &4.8 &4.9K &15 &19 \\
            
            &&prod safety&11.1 &conveyor1-3 &5 &.77/.83 &6 &13 &6 &0 &10.2 &11K &13 &24 \\
           \multirow{-2}{*}{elevator}&\multirow{-2}{*}{manufacture}& throughput&11.2  &entry conveyor &5 &.33/.68 &4 &19 &4 &0 &10 &11K &12 &24 \\
           
           \rowcolor{Gray}
           &&spacing &12.1&buffer conveyor &6 &.63/.71 &6 &13 &6 &0 &6.7 &9K &17 &13 \\
            \rowcolor{Gray}
           \multirow{-2}{*}{queue processor}&\multirow{-2}{*}{manufacture} & throughput &12.2 &entry conveyor &2 &.67/.8 &5 &11 &4 &1 &6.5 &9K &14 &13 \\
           \bottomrule
        \end{tabular}
    }
    \caption{ICS Scenarios: \sys's physical world modelling results and generated SCADA PHYSICS constraints with Diverse ICS Industry Applications}
    \vspace{-2em}
    \label{tbl:scenarios}
\end{table*}
 
\PP{Physical Process-Aware Dynamic Analysis}
\sys leverages an ICS emulation engine (FactoryIO~\cite{factoryio}) and a SCADA platform (Siemens S7 WinSPS~\cite{winsps}) to perform a \emph{physical-process-aware} dynamic analysis of process-control behavior to generate \physics constraints. FactoryIO (which also provides a SCADA tool) provides an environment to setup and drive physical processes. Its emulation engine supports hardware-in-the-loop PLCs using real protocols such as Modbus and allows loading of generic ICS scenario FBDs. 
On starting each process, we monitor the SCADA API executions to know when each process-control phase start based on the \emph{phase windows}. For each process, we \emph{induce} the SCADA execution to \emph{re-compute} the process control variable (\emph{CV}) (i.e., to send to the physical) by iteratively switching each element state in the process. This drives execution down the process-monitoring and process-altering code paths to effect change on the process, enabling \sys to record the API calls of each phase. \\
\PPPPP{Process-Aware PHYSICS Constraints}
\sys's physical model makes generated \physics constraints process-aware, i.e., captures process functions. Although many code paths exist in SCADA, our \physics constraints cover \emph{only} process-control paths by inducing the benign SCADA process-control logic to react to each element state change, ensuring that relevant "state-changing" logic paths are dynamically covered. Because element states are derived from the plant's deployed OPC, \sys's \physics constraints succinctly represent the limited legitimate APIs to control the physical.\\
\PPPPP{PHYSICS Violations by Injection} \sys detects anomalous APIs not in \physics constraints as \emph{injection} violations. As such, \sys can detect attack code injected into SCADA programs (as done by Stuxnet~\cite{stuxnet}) and redirected API calls via Import Address Table \emph{hooking}. Stuxnet-type attacks will evade existing tools that \emph{whitelists} benign SCADA programs. However, because \sys focuses on executed APIs, not the \emph{executor}, injected APIs will be detected.\\
\PPPPP{PHYSICS Violations by Bypass} Rootkits can bypass $S^{3}$ layers and directly send malicious signals to physical I/O using kernel calls such as \emph{ 0x6b DeviceIOControl}. However, because \sys sees all \emph{WRITE} traffic on the physical interface ($S0$), it detects the attack as \emph{Bypass} violation because no $S2$ activity was seen, allowing \sys to know that a kernel-space entity bypassed proper process-altering $S^3$ channels to send signals.

\section{ \xspace \xspace \xspace IMPLEMENTATION}
\label{sec:implementation}
\mr{We leveraged domain knowledge in OPC convention ~\cite{opc,opc2,opc3} to automatically extract and analyze ICS process information from input ICS scenarios files (OPC element data and FBD).}

\mr{\PP{Accessing ICS Scenario Files in plants} We developed an OPC client~\cite{opc-client} to perform $ReadRequest$ on OPC server (port 48031) using OPC UA protocol, which specifies a \emph{nodesToRead} field~\cite{opc-code-read} to return elements and parameter in JSON. Further, we read OPC Alarms\&Event data from the \emph{UaServer\_Event} structure~\cite{opc-code-event,ae} also in JSON. Lastly, we exported FBD's STL using SCADA software's provided API.}\\
\mr{\PPPPP{Process Identification and Element Tracing} 
We parse the OPC Alarm\&Event data using python to extract the monitored (i.e., terminal) element and the sensor element. The terminal element is a fitting heuristic to identify processes because it represents a process goal. Next, to identify the OPC tags in the STL and trace element's connection, we first parsed the STL and matched its tags to tags from OPC. Then, we traced the STL statements from the terminal element tag to other (non-terminal) elements. Next, our python script  “partitions” terminal and non-terminal elements in two sets.}\\
\mr{\PPPPP{Tracing SCADA APIs with LibVMI} Our testbed SCADA runs in a Windows Dom U VM in Xen Hypervisor, installed in a bare-metal intel machine, with a Linux Dom 0 VM, where \sys runs. In Dom 0, \sys's monitoring tool is implemented in C++ (77 lines) which calls LibVMI's $altp2m$ module to initialize VM introspection on our SCADA VM via $create\_view$ method. To trace SCADA execution, our C++ code invokes LibVMI's  \emph{SETUP\_INTERRUPT\_EVENT} to "trap" and forward executed APIs to a \sys's analysis tool (232 lines of python) via a Linux Pipe. We run \sys as follows:  \emph{scaphy\_monitor} \textbf{|} \emph{scaphy\_analysis.py}}

\section{ \xspace \xspace \xspace EVALUATION}
\label{sec:evaluation}
 \begin{table*}
    \centering
    \footnotesize
  \resizebox{0.94\textwidth}{!}{
        \begin{tabular}{@{}>{\columncolor[gray]{0.9}}cclll|l|>{\columncolor[gray]{0.9}}c>{\columncolor[gray]{0.9}}c|>{\columncolor[gray]{0.9}}c>{\columncolor[gray]{0.9}}c|>{\columncolor[gray]{0.9}}c>{\columncolor[gray]{0.9}}c>{\columncolor[gray]{0.9}}c|cc|c@{}}
            \toprule
            &MITRE ICS &Attack &In-the-wild &MITRE&Physical Process&\multicolumn{2}{c|}{PHYSICS}&\multicolumn{2}{c|}{Physical Anom.}&\multicolumn{3}{c|}{Signal Anom.} &\multicolumn{2}{c}{Metrics}&detect \\
            &Attack ID & Description &ICS Reference &ICS TTP&IDs (in \tabref{tbl:scenarios}))&bypass&inject&incons&setpoint&miss&extran&ooo&TP &FP&time(s)\\
            \cline{1-16}
            &T872 &wipe host/registers &Killdisk &Evasion &1.1, 1.2, 4.1                                &\cmark &           &\cmark       &       &    &\cmark      &          &3       &0&9.9\\
           &T836 &modify parameter &Stuxnet &Impair proc contrl&4.1, 4.2, 8.1, 8.2              &       &\cmark     &       &\cmark       &    &\cmark      &       &3       &0&9.0\\
            &T831 &contrl manipulation &Stuxnet &Impact control&2.1, 2.2, 6.2                 &       &\cmark     &       &       &\cmark     &      &\cmark        & 3      &1&10.2\\
            &T889 &kernel driver attack &Blaster &Modify Program fxn&5.1, 5.2, 10.2               &\cmark &           &       &\cmark       &    &      &          &2       &0&9.8\\
            &T855 &unauthorised cmd msg &Industroyer &Impair proc contrl&3.1, 3.2, 11.1    &       &\cmark     &\cmark       &       &    &\cmark     &       &3       &1&7.8\\
           \multirow{-6}{*}{\rotatebox{90}{Process} \rotatebox{90}{Altering}}  &M1.2 &corrupt registers &Triton &Impair proc ctl &7.1, 10.2, 11.2    &       &\cmark     &\cmark 
           &       &    &\cmark      &\cmark      &4       &0&8.8\\
            \cline{1-16}
            &T874 &library hooking &Triton &Execution&5.1, 5.2, 10.1, 10.2                              &       &\cmark     &       &       &    &      &          &1       &0&5.4\\
            &T801 &monitor proc state &Industroyer &Collection&6.1,6.2, 8.1                               &        &\cmark    &       &       &    &      &         &1      &0&4.1\\
            &T861 &points/tags identifica. &Backdoor.Oldrea &Collection (OPC)&6.1, 6.2, 8.2                &       &\cmark     &       &       &    &      &       &1       &0&4.1\\
            &T816 &device shutdown &Industroyer &Inhibit Resp fxn&7.1, 9.1, 9.2, 7.2            &       &\cmark     &       &       &    &      & &1           &0&5.4\\
            \rowcolor{Gray}
            &T888 &network Enumeration &Havex(as is) &Discovery fxn&4.1, 4.2, 8.1, 8.2            &       &\cmark     &       &       &    &      & &1           &0 &4.8\\
            \rowcolor{Gray}
            \multirow{-5}{*}{\rotatebox{90}{Non-Process} \rotatebox{90}{Altering}}&T805 &block serial COM &Industroyer(as is) &Inhibit Resp fxn&1.1, 1.2, 9.1             &       &\cmark     &       &       &    &      &         &1       &0 &5.1\\
            \bottomrule
        \end{tabular}
        }
    \caption{Deployed Attacks and Detection Metrics. We leverage the MITRE ICS Attack Framework~\cite{mitre} to categorize the attack TTPs }
    \label{tbl:attacks1}
    \vspace{-3em}
\end{table*}

We evaluate \sys's ability to (i) detect a variety of ICS attacks across diverse ICS scenarios, and (ii) outperform existing tools in detection accuracy.
We launched \attacks ICS attacks on \scenarios diverse ICS scenarios across \industries ICS industries to show \sys versatility, as shown in \tabref{tbl:scenarios}. \sys detected \accuracy of all attacks, including real ICS malware (Havex\footnote{7f249736efc0c31c44e96fb72c1efcc028857ac7} and Industroyer\footnote{2cb8230281b86fa944d3043ae906016c8b5984d9}), with only \fp false positives.
Due to lack of resources to support diverse ICS security research~\cite{swat2,swat1, morris1, morris2}, we make available over 200GB of new ICS experiments and attacks in both SCADA and physical aspects, developed for FactoryIO Engine~\cite{factoryio} and Siemens S7 platform~\cite{winsps}.\\
\PPPPP{Experimental Setup}
\label{ssec:labsetup}
We leveraged a U.S. national lab testbed, which supports fast deployment of ICS topologies, OPC, HMIs, and SCADA VMs, prepared with ICS tools to control physical processes such as UART interfaces and Windows Serial Driver. We used 3 SCADA platforms: S7 WinSPS, MyScada~\cite{myscada}, and FactoryIO SDK. This makes our testbed suited to evaluate \sys against ICS attacks in realistic settings. To support diverse processes, we leveraged Simulink~\cite{simulink}, PowerWorld~\cite{powerworld}, and FactoryIO Engine to emulate physical processes in Remote Terminal Units (RTUs). \\
\PPPPP{ICS Attacks Performed}
We performed diverse modern ICS attacks from 4 categories: attacks that (\textbf{I}) maliciously alter element states in running processes, (\textbf{II}) blocks SCADA access to the physical, (\textbf{III}) collects attack-relevant data from SCADA, and (\textbf{IV}) exploit bugs in ICS devices. We leveraged Mitre ICS Attack Framework and ICSSploit~\cite{icssploit} to develop realistic attacks, indicated in "In-the-wild" column of \tabref{tbl:attacks1} and \tabref{tbl:attacks2}, each tailored against their pertinent ICS target.

\PP{Physical World Model and PHYSICS Constraints}
\tabref{tbl:scenarios} shows our ICS scenarios details and \sys's derived physical world models and \physics constraints. To verify accuracy of generated PHYSICS constraints (API calls), we produced forensic execution traces of our SCADA program using the Time-Travel debugging feature of Windows Debugger (WinDBG), which we manually stepped to see the APIs of each process-control path.
\tabref{tbl:scenarios} column 8-11 shows the number of unique \alter APIs called on average per path. As shown, we found that only few unique APIs were seen (most times in loops) depicting that \alter is very specialized per physical domain, but the high stack depth shows that \alter is executed deep in SCADA logic. We found that the FPs were due to rare element states not parsed correctly from OPC.
\begin{table*}
    \centering
    \footnotesize
  \resizebox{0.87\textwidth}{!}{
        \begin{tabular}{@{}lllll|>{\columncolor[gray]{0.9}}c>{\columncolor[gray]{0.9}}c>{\columncolor[gray]{0.9}}c>{\columncolor[gray]{0.9}}c>{\columncolor[gray]{0.9}}c>{\columncolor[gray]{0.9}}c>{\columncolor[gray]{0.9}}c|ccc@{}}
            \toprule
            ICSSPLOIT&Attack &Real-world&In-the-Wild &Exploit&\multicolumn{7}{c|}{\emph{SCADA Software Stack} ($S^3$) Activity}  &\multicolumn{3}{c}{\emph{Metrics}}\\
            Attack ID & Description &Device Targets & CVEs/ICSAs& Type&$S4.1$&$S4.2$&$S4.3$&$S3$&$S2$&$S1$&$S0$&TP&FP&FN\\
             \cline{1-15}
            SPLOIT1.1 &stop controller/CPU &Siemens Simatic-1200 & ICSA-11-186-01&Unprotected Port&-&-&\cmark&\cmark&\cmark&\cmark&\cmark&5&0&2\\
            SPLOIT1.2 &remote code execution &QNX SDP 660  &CVE-2006-062&Buffer Overflow&-&\cmark&\cmark&\cmark&\cmark&\cmark&\cmark&4&0&1\\
            SPLOIT1.3 &remote device halt &Schneider Quantum &ICSA-13-077-01&I/O corruption&-&\cmark&\cmark&\cmark&\cmark&\cmark&\cmark&5&0&1\\
            SPLOIT1.4 &crash RTOS service &QNX INETDd&CVE-2013-2687&Buffer Overflow&-&\cmark&\cmark&\cmark&\cmark&\cmark&\cmark&5&0&1\\
            SPLOIT1.5 &RPC device crash &WindRiver VXWorks& CVE-2015-7599&Integer Overflow&-&\cmark&\cmark&\cmark&\cmark&\cmark&\cmark&4&0&1\\
            SPLOIT1.6 &denial of service &Siemens S7-300/400& CVE-2016-9158&Input Validation&-&-&\cmark&\cmark&\cmark&\cmark&\cmark&4&0&2\\
            \bottomrule
        \end{tabular}
        }
    \caption{Deployed Attacks and Detection Metrics. We leverage Attack Modules ICSSPloit Attack Modules~\cite{icssploit} to categorize the attack TTPs }
        \vspace{-3em}
    \label{tbl:attacks2}
\end{table*}
\sys's physical model allowed it to prune non-impactful elements from the original pool extracted from OPC, which is efficient. E.g., many ICS elements such as repeaters do not have any impact on process output, hence were pruned off during \ic derivation. As such, we saw over 50\% reduction (on avg) from extracted OPC elements to PDIG nodes. This outperforms naively analyzing all states regardless of impact as done in~\cite{state}.

Further, the reduced element pool contributed to an \ic (impact) average above 70\% of their process output, showing that \sys modelled the relevant or \emph{impactful} elements whose malicious state change are more disruptive to the process. \sys physical-process-aware dynamic analysis takes about 8 mins, which is reasonable per deployed scenarios sizes (last 3 columns). Our ICS scenarios are adapted from real-world models developed. For example, our power grid scenario was adapted from an open-source Texas Pan Handle power grid~\cite{texas}, simulated in \emph{PowerWorld} RTUs. We use the power grid example to explain \tabref{tbl:scenarios}: \sys accurately identified the process terminal element ($E_{Term}$) as shown. \sys's impact analysis pruned the OPC element pool of 19 nodes to 10 PDIG nodes, which is efficient. The average impact of all PDG nodes was over 50\% with Max at 0.76 and 0.66, meaning that attack involving them are most disruptive.

\subsection{ICS Attack Detection}
\label{ssec:attacks}
\tabref{tbl:attacks1} shows attack categories \textbf{I}, \textbf{II}, \& \textbf{III}, and \sys's results. Attack category I are shown in the first row-group, "Process Altering". Category II \& III are shown in the second-row group, Non-Process Altering. \sys detected \physics bypass and inject attacks for both categories. However, \sys detected physical and signal anomalies for only category I, because category II \& III do not send control signals but block SCADA access to the physical or collect data about devices. For example, in $T805$, Industroyer issued \emph{CreateFile} on all \emph{COM} ports to block our SCADA program from accessing the physical ("as is" mean we executed In-the-wild malware). Similarly, Havex in $T888$ enumerated all \emph{COM} device objects stored in Registry keys \emph{HKLMSYSTEMCurrentControlSetServices} (registry value \emph{SERVICE\_KERNEL\_DRIVER} are device driver services) to identify connected devices by issuing loops of \emph{OueryKey}, \emph{OpenKey}, \emph{QueryValueKey} API calls. These API calls were atypical of process-monitoring, allowing \sys to detect their \physics inject violations.

\sys detected two \physics bypass violations in category I, $T872$ \& $T889$, which are kernel attacks that bypassed \sys monitored $S^3$ layers. Specifically, $T889$ \& $T872$ used \emph{DeviceIOControl} direct driver call to send signals out the serial I/O. \sys detected the attack because they sent out signals without accessing $S^3$ layers, allowing \sys to know that a kernel entity bypassed proper \alter $S^3$ channel to attack the physical.
Further, $T872$ used the same call to send control code to the storage device driver to delete whole drives (/DosDevices/C:). The Last 3 columns show results. Ground truth was derived from open-source attack data~\cite{mitre, metasploit}. We found that FPs were due to missed APIs during \physics analysis due to rare element states not parsed from OPC.

\PP{Evaluating SCADA Software Stack ($S^3$) Activity}
For attack category \textbf{IV}, we demonstrate $S^3$'s ability to \emph{pinpoint} steps along SCADA access to the physical, which shows that $S^3$ layers are practical host \emph{monitoring taps} for ICS attack detection. To do this, \sys tracked access to each $S^3$ layer based on the API call identifiers for each layer. E.g., \emph{ReadFile} accesses $S2$, the ICS device object. We leveraged ICSSPLOIT~\cite{icssploit} to compile real-world exploits that were developed as proof-of-concepts exploit code against real bugs in ICS devices. We launched these attacks against their simulated ICS targets and then check which $S^3$ component was accessed in the SCADA host. \tabref{tbl:attacks2} details the results. \sys detected all five layers of $S^3$ in all attacks as shown. However, in layer $S4$, \sys did not detect \init behavior ($S4.1$). This is because the exploits where self-contained and did not issue any API to setup environment. However, \sys detected their \alter behavior ($S4.3$) when the \emph{WriteFile} API was called to send signal to the physical.
We hope that via $S^3$, Antivirus companies can develop SCADA-specific host agents to correlate accesses to specific $S^3$ layers to detect attacks.

\subsection{Case Study: 2021 Florida Water Plant Poisoning Attack}
\label{ssec:florida}
\begin{figure}[t]
\minipage{0.24\textwidth}%
\begin{subfigure}[c]{\textwidth}
  \includegraphics[width=0.93\textwidth]{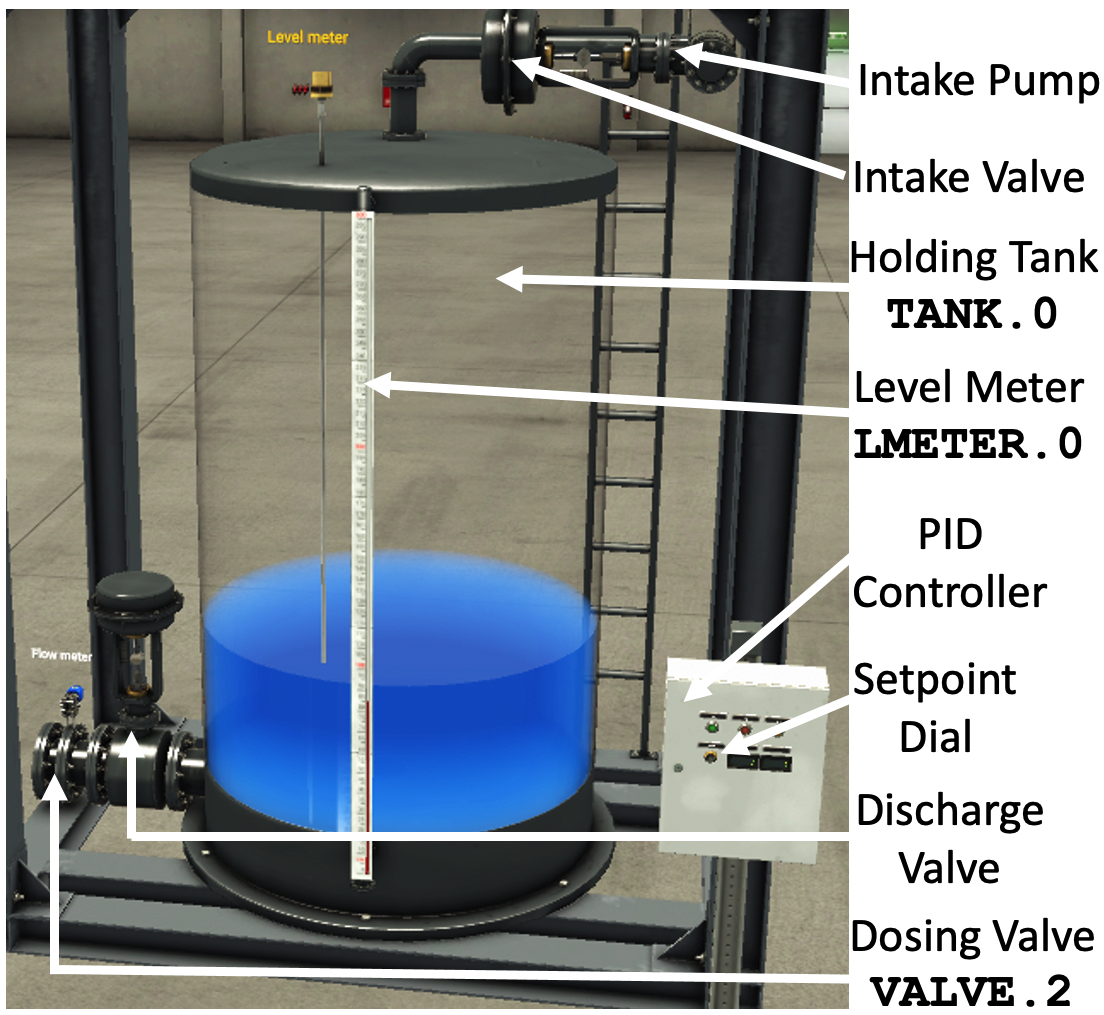}
  \subcaption{Chemical Dosing Operation HMI}
  \label{fig:dosing}
  \includegraphics[width=1.0\textwidth]{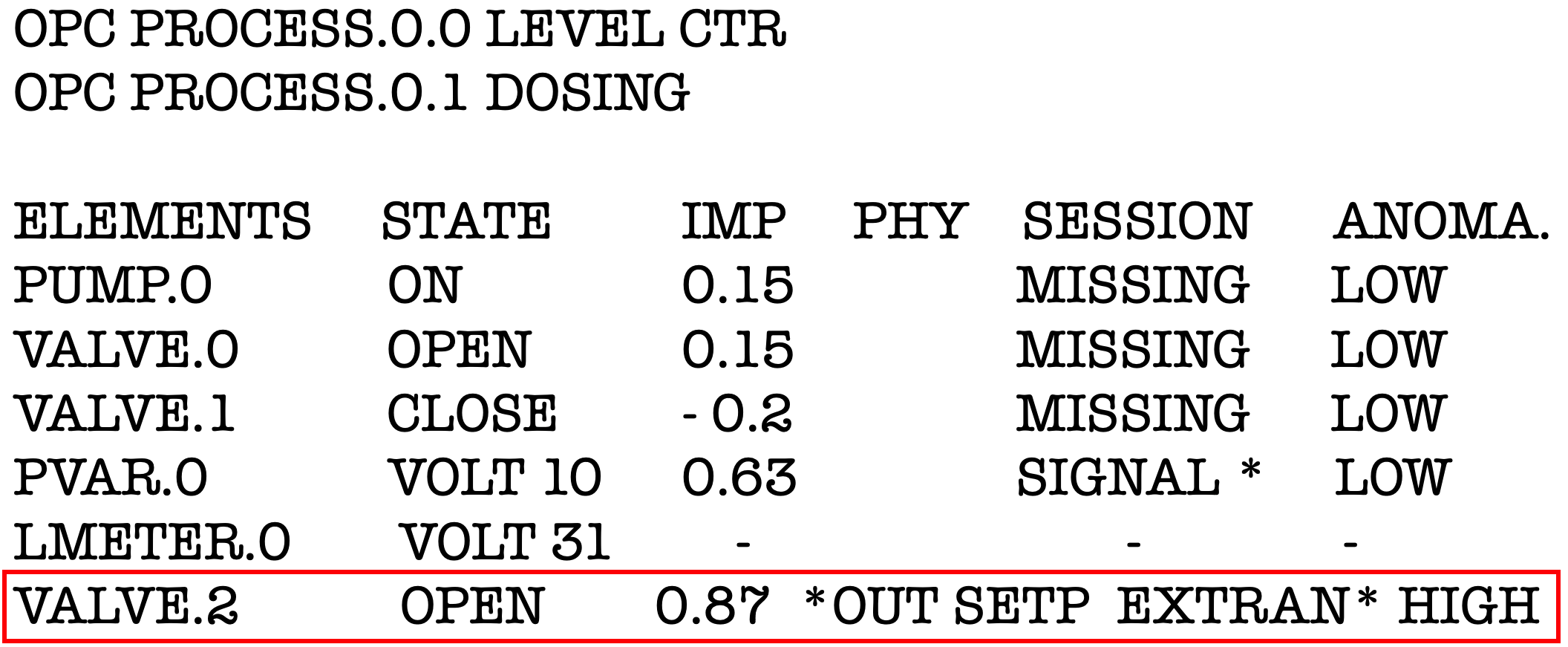}
  \caption{\sys Output}
\label{fig:florida_output}
\end{subfigure}
\endminipage
\minipage{0.20\textwidth}%
\begin{subfigure}[c]{\textwidth}
  \includegraphics[width=0.99\textwidth]{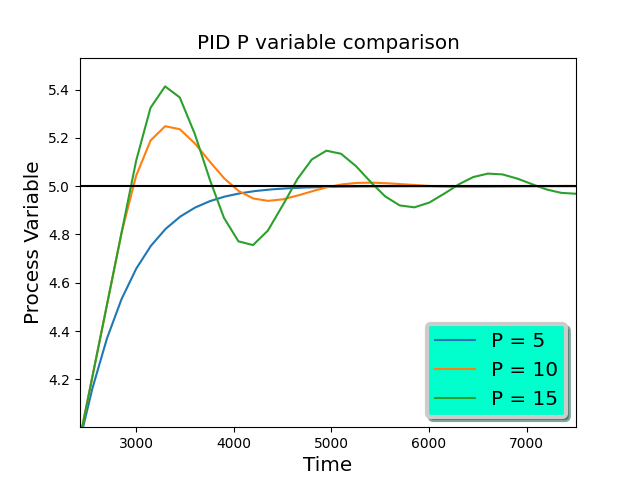}
  \caption{Effects of different P's}
  \label{fig:pchanges}
  \includegraphics[width=0.99\textwidth]{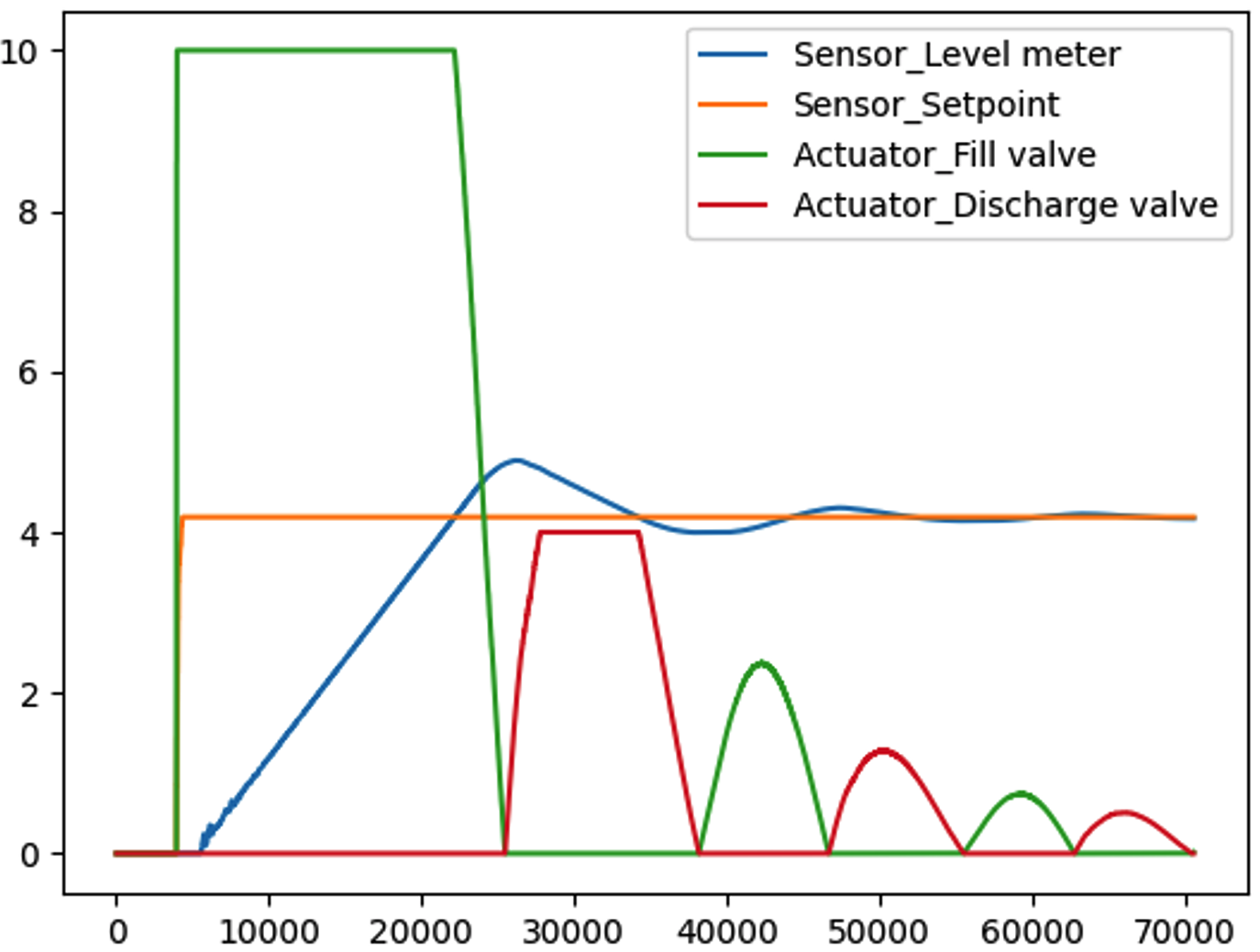}
  \caption{Intake/discharge cycles}
  \label{fig:pid1}
 \end{subfigure}
\endminipage
\vspace{-1em}
\end{figure}

We replicated the FL incident with realistic water treatment scenario using FactoryIO and open-source data~\cite{swat1, swat2}. The attacker targeted the chemical dosing operation by manipulating a \emph{proportional} $P$ parameter to raise toxic levels of NaOH in the water outside the setpoint (\emph{SV})~\cite{florida}. Chemical dosing involves two processes: \emph{Level control} (LC) and \emph{Dosing}. The HMI is shown in \fref{fig:dosing}. LC aims to fill a holding tank, \emph{TANK.O} with chemical based on \emph{SV}, after which Dosing will open a valve, \emph{VALVE.2} to let chemical into the water supply~\cite{swat1, swat2}. LC is controlled by a physical domain logic, \emph{Proportional Integral Derivative} (\emph{PID}). Because \emph{SV} cannot be reached in one shot, PID operation involves several "intake" and "discharge" cycles, (shown in \fref{fig:pid1}) whereby an intake pump fills chemical into \emph{TANK.O}, and a discharge valve remove accesses.
$P$ controls how aggressive the intake and discharge cycles are driven. E.g, a high $P$ pumps an \emph{initial} excessive volume into \emph{TANK.O}. \fref{fig:pchanges} shows how different $P$ values affect how \emph{SV} is reached. \emph{SV} is set via a hardware dial on the PID controller, so attacker cannot modify it via cyber.
\\
\PP{Attack and Detection} The attacker issued 2 control signals to raise $P$ (dumps excess chemical into \emph{TANK.O}) and open \emph{VALVE.2}. We launched the attack using a Modbus payload, but made the attack code self-contained, without triggering any \physics violation. At the SCADA side, the attack triggered all $S^3$ layers to send out signal, which was not malicious by itself. We now focus on the physical aspects. \sys's detected a high \emph{outside set point} physical anomaly and a low \emph{Extraneous} signal anomaly. \fref{fig:florida_output} shows \sys's outputs.\\
\textbf{Explanation} \sys detection is based on the functional process relationship captured in the physical model between LC and Dosing. Recall \ssecref{ssec:physical}, \emph{VALVE.2} is a PTP element between LC (PTP source) and Dosing (PTP sink). When \emph{VALVE.2}'s state changes from \emph{CLOSE} to \emph{OPEN}, the Dosing process output \emph{assumes} \emph{TANK.O}'s value (i.e., the $E_{Term}$ of the PTP Source) which is measured by the meter sensor \emph{LMETER.0}. This Dosing process outcome (\emph{LMETER.0}'s value) was outside the setpoint for Dosing derived during \ic derivation, which allow \sys to detect it. Finally, a high anomaly score is calculated based on \emph{VALVE.2}'s \ic of 0.87, allowing \sys to detect an \emph{outside setpoint} anomaly. Further, \sys detects an \emph{extraneous} signal anomaly because the signal to open \emph{VALVE.2} was "extraneous" in LC.
\subsection{Case Study: Reliability of \sys against Modern Rootkits that knows \sys Approach}
\label{ssec:casestudy}
\mr{To test \sys's reliability against rootkits, we selected two modern rootkit techniques based on recent works~\cite{rootkits1, rootkits2, rootkits3}: (i) direct kernel object manipulation (DKOM), which tampers kernel objects, and (ii) Hypervisor-level DKOM. We show that \sys can be reliable against DKOM rootkits, but not hypervisor-level DKOM. We leveraged Metasploit~\cite{metasploit} to develop these rootkits. We note that \sys only applies to rootkits that \emph{attack the physical world from SCADA} (i.e., SCADA rootkits). That is, \sys is not aimed to detect traditional IT-only rootkits (e.g., exploit attacks or backdoors). \sys's ability to detect \emph{evasion} in SCADA is because any SCADA entity must ultimately send out traffic via the physical interface to access the physical. To do that legitimately, they must traverse proper SCADA's $S^3$ channels (monitored by \sys). If a rootkit naively \emph{circumvents} these channels, but sends signals via the physical interface, \sys detects the evasion because no $S^3$ API calls was seen, which means a kernel entity (e.g., rootkits) bypassed proper $S^3$ channels. This was the case in \ssecref{ssec:attacks}, where rootkits $T872$ \& $T889$ used kernel call \emph{DeviceIOControl} to send signals, without $S^3$ activities. Hence the goal of a SCADA rootkit in evading \sys is to deceptively "present" \sys with proper $S^3$ API behavior, while sending commands to the physical.}

\mr{\PP{DKOM Rootkit} This rootkit installs as a kernel driver and uses DKOM to obtain (and execute on) the userspace ICS device object ($S2$ layer). Unlike naive rootkits $T872$ \& $T889$, this rootkit exploits \emph{ZwCreateFile} native API which handles parameters differently in kernel than the userspace \emph{CreateFile}~\cite{zw}. Calling \emph{ZwCreateFile} with the driver name returns the driver base object address (a kernel object) which is traversed to locate \emph{ICS device objects} in driver data structure. With this object, the rootkit (i) inserts itself on the ICS Device Stack of the physical interface, from which it can send signals, and (ii) invokes \emph{ZwWriteFile} to complete fake $S^3$ behavior.}\\
\begin{figure}[h]
\centering
  \vspace{1em}
\includegraphics[width=0.40\textwidth]{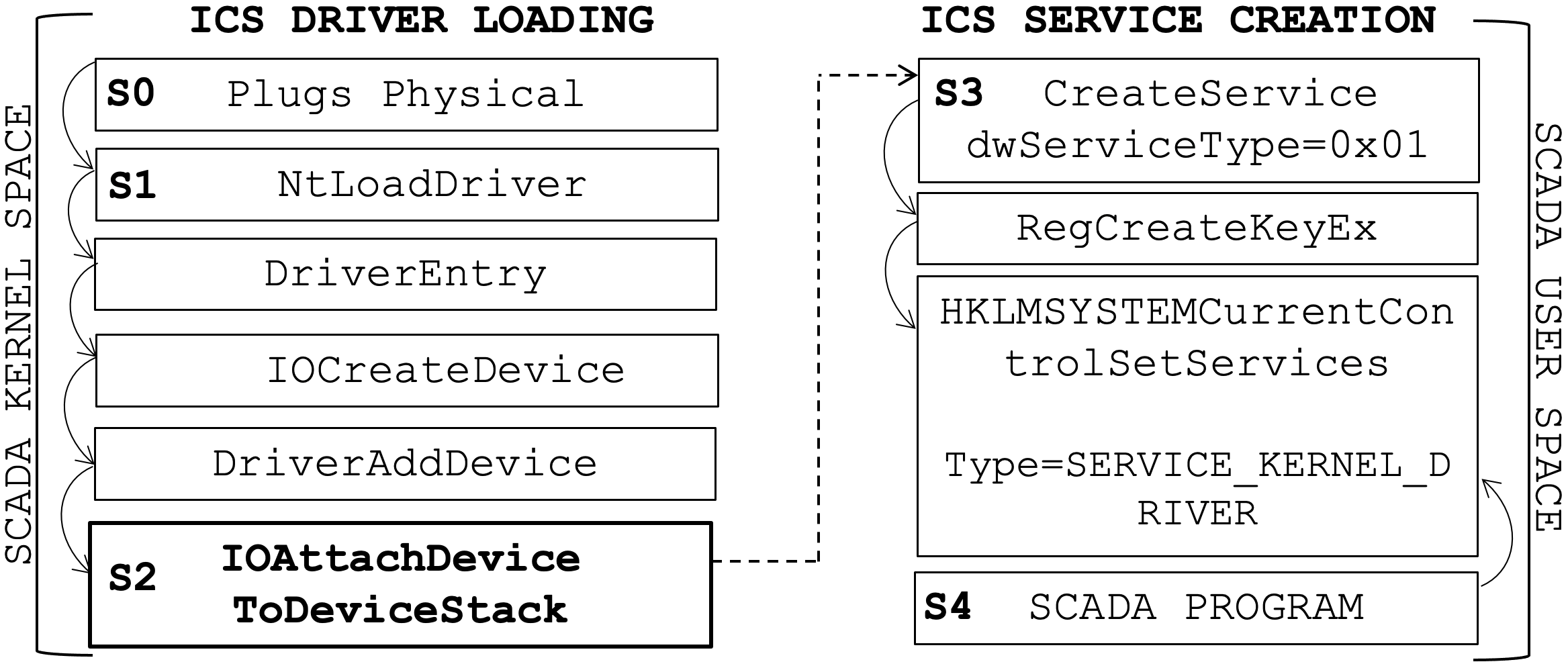}
  \caption{API sequence for ICS Driver Loading and Registration. }
\label{fig:init}
\end{figure}
\mr{\PP{Detection} Executing the rootkit emitted $S2$ \alter APIs with no \physics violations based on \emph{ZwCreateFile} \& \emph{ZwWriteFile} calls. This evades \sys. However, \sys can detect this rootkit by simply monitoring for ICS driver attachment to the Device Stack, a protected kernel object.}\\
\mr{\textbf{Explanation} For the attack to work, the rootkit had to \emph{attach} itself on the \emph{ICS Device Stack} of the physical interface. This is done using \emph{IoAttachDeviceToDeviceStack} during driver loading as shown in
\fref{fig:init}, which shows the steps ICS drivers (and rootkit) must follow to attach themselves to a Device Stack. Device Stack is a kernel structure that tells the kernel which drivers can access I/O of a physical device. Hence \sys can track when new drivers are added to a Device Stack, allowing it to catch when \emph{additional or unknown} drivers are introduced.}

We note that rootkits cannot modify the kernel subsystem (\emph{NTOSKRNL.EXE}) per \emph{Windows Kernel Patch Protection (KPP)}~\cite{kpp}, which prevents third-party code (e.g., drivers) from "patching" the kernel subsystem, which includes Device Stack. \mr{KPP prevents unauthorized access of device I/O. Hence ICS drivers (and rootkits) must follow the API steps in \fref{fig:init}, which \sys can monitor with high-fidelity.
To test this, we added this API behavior to \sys model and ran the rootkit. On observing a repeat of \emph{IoAttachDeviceToDeviceStack} (i.e., \emph{after} the benign ICS driver has already been attached), \sys detects this malicious Device Stack tampering. Making this addition to \sys was seemless because the API sequence in \fref{fig:init} \emph{follows} \sys's proposed $S^3$ layers, which shows the ubiquitousness of the $S^3$ reference model. For example, the argument \emph{dwServiceType}=$0x01$ in \emph{CreateService} registers an ICS driver service, which returns a handle used by ICS callback functions ($S3$) to access device objects ($S2$).}

\mr{\PP{Hypervisor-Level DKOM} This second attack requires the rootkit to \emph{escape} the VM and launch DKOM in Xen hypervisor (Dom 0 Linux) where \sys runs. Although this is outside our threat model, as the hypervisor is our TCB (\secref{sec:threat_model}), we wanted to show a concrete way to evade \sys. Recall in \secref{sec:implementation} that \sys's LibVMI-based monitoring tool uses a Linux Pipe to feed API traces to \sys analysis tool. This DKOM works by redirecting this Pipe's memory buffer in kernel to another pipe the rootkit controls. Specifically, our rootkit modified the Page Pointer, \emph{PAGE *}, the first argument of \emph{pipe\_buffer}, a struct member of \emph{pipe\_inode\_info} kernel object, which manages Pipe operations in Linux kernel. Since Linux Pipes are mounted in file-system "\emph{pipe:/}" it was easy to find \sys's pipe via \emph{ls}. Then, the rootkit modified \sys's Pipe's \emph{PAGE *} in kernel to a malicious Pipe buffer address created with \emph{dcom\_write} \textbf{|} \emph{dcom\_read}, which allowed attacker to control what the \sys analysis sees, without \sys knowing. In our attack, we killed the LibVMI side of \sys after the pipe manipulation for the attack to work.}

\begin{table}
    \centering
    \footnotesize
\resizebox{0.96\columnwidth}{!}{
        \begin{tabular}{@{}l|l|c|ccc|ccc|ccc@{}}
            \toprule
             &  
             &&\multicolumn{9}{c}{\textbf{Attack Detection Metrics}}\\
             \textbf{Techniques} 
             & Approach 
             &Attacks/&\multicolumn{3}{c|}{SCADA}&\multicolumn{3}{c|}{Physical}&\multicolumn{3}{c}{CTRL Signals}\\
                                          &
                                          &Normal      &TP    &FP    &FN     &TP   &FP      &FN     &TP     &FP  &FN\\
             \cline{1-12}
             Sensor~\cite{state}& Linear    
             &40/146&-     &-     &-      &19    &37     &21       &-        &-        &-\\
             Analysis &Regressive M. 
             &&&&&&&&&&\\
             
             \cline{1-11}
             \rowcolor{Gray}
             Traffic~\cite{telemetry}  &  Decision Tree
             &40/146&-     &-     &-  &-     &-     &-  &11    &18     &29 \\
             \rowcolor{Gray}
              Analysis & Classifier  
              &&&&&&&&&&\\
            \cline{1-12}
             Hybrid & SCADA Corr.
             &40/146   &36  &5     &4         &21  &14   &19        &18   &21   &22      \\

             \sys &with Physical 
             &&&&&&&&&&\\
           \bottomrule
        \end{tabular}
        }
    \caption{Comparison with Existing Techniques}
    \vspace{-1em}
    \label{tbl:comparison}
\end{table}

\subsection{Comparing \sys with Existing Techniques}
\label{ssec:compare}
We compared \sys to existing ICS techniques that use physical models~\cite{state} and traffic classifier~\cite{telemetry}.~\cite{state} is based on~\cite{limiting} and analyzes sensor data's cumulative sum of residuals, and raises alarm if the difference between sensor and expected behavior is higher than a threshold.~\cite{telemetry} analyzes spatial-temporal properties of ICS signals such as packet arrival times and size, using a REBTree classifier.~\cite{telemetry} raises alarm when traffic features are \emph{outside} a running average. To do this comparison, we use sensor and traffic data from the experiments in \tabref{tbl:scenarios}, which we parsed into \emph{@.ARFF} format \mr{(sample shown in \secref{sec:appendix}), and make available via this work} We leveraged open-source tools to setup these techniques. For~\cite{state}, we leveraged Scikit-Learn to generate a linear regressive model to fit the sensors values of the physical elements in the normal running mode. For~\cite{telemetry}, we leveraged WEKA~\cite{weka} to generates a REPTree classifier that builds a decision tree using information gain and variance in the extracted traffic fields.

To launch attacks, we follow the format in \tabref{tbl:attacks1}, which produced 40 attacks and 146 normal instances. \tabref{tbl:comparison} shows the results. Existing tools did not detect any SCADA attacks due to no SCADA context. However, this is where \sys detected most attacks (90\%, and \accuracy overall). \mr{We note that diverse ICS experiments (such as ours) may affect the performance of tools designed for specific ICS domains}.~\cite{state} detected 19 attacks (\existaccuracy), with a high FP of 37 (\existfp). Its FP is due to flagging high sensor deviations that are part of benign behaviors. One instance is the FL water attack where a high $P$ variable causes a high but \emph{temporary} sensor deviation. Although it deviates greatly from the \emph{setpoint}, it is benign in Level Control (LC), but anomalous in the Dosing process if dosing valve is open. Unlike \sys's physical model, existing linear models~\cite{state,limiting,ssa,ar} do not capture inter-process operations (e.g., between Dosing and LC), hence are prone to false alarms due to high but temporary benign deviations. \sys's physical model lower TP is due to approximating analog states such as switching 0-10v with only 3 levels 0v,5v,10v, which saves space but is less precise.~\cite{telemetry} detected 11 attacks (\secondexistaccuracy), but with low FP (\secondexistfp) because most modern attack is similar to benign. \sys's signal-based detection had more FPs because it flagged many benign missing signals, showing that missing signals are not effective, which we can mitigate by raising its detection threshold.

\subsection{Quantifying Manual Work to use SCAPHY in Practice}
\label{ssec:manual}
\mr{We quantify the amount of manual work that may be required to apply \sys's pipeline in practice. For example, although our SCADA platform provides interfaces to automatically export FBDs to STL, other platforms may require manual export via the GUI. In addition, automatically matching and tracing OPC tags in the STL require SCADA and OPC to use the same namespace (e.g., "\emph{BRK}" in OPC is also "\emph{BRK}" in STL). This was the case in our testbed and in integrated platforms such as Siemen's TIA. However, in plants using third-party tools such that STL and OPC have different namespace (e.g., "BRK" v. "BR"), an STL-to-OPC element matching can be achieved semi-automatically by manually seeding a regular expression script to perform the matching. \tabref{tbl:manual} summaries the time to perform each step and repetitions needed.} 
\begin{table}
    \centering
    \footnotesize
\resizebox{0.82\columnwidth}{!}{
        \begin{tabular}{@{}ccc|>{\columncolor[gray]{0.9}}c|>{\columncolor[gray]{0.9}}c@{}}
            \toprule
             \mr{\textbf{\sys Pipeline} }&\mr{ Where }&\mr{ Manual }&\mr{Avg. }&\mr{ Times }\\
            \mr{ \textbf{Requirement}}&\mr{Used }&\mr{Step}&\mr{ time }&\mr{repeated }\\
             \cline{1-5}
            \mr{ Exporting FBD }&\mr{ Element }&\mr{Several Clicks in}&\mr{under 1 }&\mr{1 per }\\
             \mr{to STL JSON }&\mr{ Tracing }&\mr{SCADA GUI }&\mr{minutes}&\mr{ scenario }\\
              \cline{1-5}
              \mr{Matching STL}&\mr{ Element }&\mr{Seeding a Regex script}&\mr{under 10 }&\mr{1 per}\\
              \mr{to OPC tags }&\mr{ Tracing }&\mr{with match parameter}&\mr{minutes}&\mr{scenario}\\
               \cline{1-5}
               \mr{Setting up SCADA}&\mr{ Impact }&\mr{Configuring testbed}&\mr{under 1 }&\mr{\mr{1 per}}\\
              \mr{Modelling ENV }&\mr{ Derivation }&\mr{networking \& devices }&\mr{hour}&\mr{scenario}\\

           \bottomrule
        \end{tabular}
        }
    \caption{\mr{Quantifying amount of manual work to apply \sys in practice}}
    \vspace{-1em}
    \label{tbl:manual}
\end{table}

\section{\xspace \xspace \xspace \xspace \xspace \xspace Discussions: \sys's Practicality}
\label{sec:discussion}
\PPPPP{Runtime Overhead}
\sys leverages LibVMI state-of-the-art VM analysis tool in Xen Baremetal hypervisor, which enables detection of malicious APIs in SCADA VMs around 9 seconds, as shown in \ssecref{ssec:attacks}. LibVMI achieves near-native speed access of guest VM memory pages when run on Baremetal Hypervisors, and now used in production~\cite{fastvmi2, fastvmi3, fastvmi, fastvmi4, fastvmi5,drakvuf}. This makes \sys a practical detection technique. In contrast, existing work based on sensor data~\cite{state, limiting}, must observe several sensor deviations before making a \emph{sound} decision, which pushes their \emph{time-to-detect} to the minute range~\cite{state,limiting}.\\
\mr{\PPPPP{Real vs. Emulated State Switching} As switching real processes can be dangerous, we used emulated processes. However, the switched states were informed by deployed OPC server in the  plant. Real-world gaps exist if poor emulation is used. To reduce this practicality gap,  SCAPHY leveraged state-of-the-art Simulink and PowerWorld ICS emulation.}\\
\PPPPP{ICS Attack Dev. Difficulty} 
Unlike IT, developing ICS attacks is hard due to finding their SCADA and physical target. As such, we spent months developing many modern attack scenarios, and tested more attacks than existing work. \\
\PPPPP{Robustness of PHYSICS Constraints}
\physics constraints are generated per ICS scenario and SCADA. This is practical because plants rarely update physical domain logic, since they are based on immutable laws of physics. 

\PP{Limitations and Future Work}
\sys cannot detect attacks that originate outside of SCADA such as side channels and device hardware. We will investigate similar execution phase-based analysis these ICS threat models.
Further, although \sys can rely on Window's KPP to detect rootkits evasion, with time new rootkits may bypass KPP. We will investigate SCADA-specific defenses for rootkit such as integrating Call Stack analysis. This shows promise given that SCADA \control have well-defined Call Stack behavior (\fref{fig:wave}) which can be robust (but expensive) than API calls.

\section{ \xspace \xspace \xspace RELATED WORKS}
\label{sec:related_works}
What differentiates ICS from IT is that physical tasks follow immutable laws of physics~\cite{limiting,state}, which can be learned to build prediction models~\cite{learn-physics1}. Existing work build sensor behavior models~\cite{ar, state, limiting, ssa} to predict when behaviors deviate from expected. However, in practice, such models are not always available~\cite{ssa, debunk}, and raise false alarms due to noise and config changes~\cite{ssa}. 
Offensive ways such as Harvey~\cite{harvey} can MITM sensors and present "false" data.~\cite{mitm1,mitm2,mitm3} proposed ways to address MITM.~\cite{diode} uses non-PLC diode gateways to avoid MITM. \mr{Reinforcement and Deep Learning~\cite{rl1,rl2,rl3,rl4} uses game-theory to learn normal and attack behaviors, but requires a high-interaction system, known attacks, and expert reward function, which may limit its use in diverse ICS practice.}

Statistical analysis of ICS traffic~\cite{dnp_attack, celine, telemetry, modbusmodel, justtraffic, stats1, ml, topologychanges,ocsvm1} are effective for \emph{noisy} and abnormal traffic such as illegal protocols and scans. However, modern attacks evade them using benign protocols and knowledge of parameters to cause specific (not noisy) attacks~\cite{industroyer, industroyer2, sok}. Flow-based approaches~\cite{dnp_attack, justtraffic, modbusmodel, functioncode} analyze abnormal function codes/channels, such as shown in~\cite{attack1, attack2}, but are evaded by attacks such as Industroyer, which uses legitimate HMIs. Timing analysis~\cite{celine,timedetection,telemetry} analyze anomalous round trip time delays and inter-arrival times. However, they are effective for signals that are \emph{chatty}~\cite{celine} such as attacker scans, not modern attacks that are targeted. \ \mr{Lee~\cite{lee} only monitors for host DLL injection, which may not happen in ICS attacks. In contrast, \sys deems API calls as anomalous when executed in the wrong SCADA execution phase.} \mr{Side channel defenses such as EMF and power~\cite{sc-emf, sc-power} may require close proximity (motherboard-level) to controllers.}

Pattern-based analysis~\cite{pattern1, pattern2, modbusmodel} detects anomalous signals such as isolated signals but suffers from model's high sensitivity, which causes false alarms due to slight config changes (e.g., addition of new devices)~\cite{modbusmodel, pattern1}. State-based tools~\cite{pattern1} detects \emph{critical states} in ICS but requires manual rules, which do not scale. Further, because they analyze all state transitions, can suffer from state explosion when parameters increase. \sys is stateless and only analyze current and \emph{impactful} element states. \mr{Process-aware tools~\cite{bro, runtime-powerflow,process-semantics,process-awareness,runtime-monitoring,ar,ssa} analyze sensor data unique to specific process functions, which may reduce ambiguity in detection, but requires experts to specify safety threshold violations per ICS-domain. This results in a tradeoff of being manual or domain specific.
For example, ~\cite{process-semantics, bro} uses manual BRO rules (e.g., heat level must not exceed 20). ~\cite{runtime-powerflow} uses expert-software to predict power flows. ~\cite{process-awareness} uses pre-defined power flow rules to analyze deviations in measurements. In contrast, \sys's process-aware physical model automatically detects physical anomalies by analyzing the disruptive "impact" of control signals. Further, unlike the above techniques, \sys correlates physical "impact" with behaviors in SCADA (i.e., anomalous API calls in atypical execution phase) for contextual "end-to-end" attack detection.} 
\section{\xspace \xspace \xspace \xspace CONCLUSION}
\vspace{-.5em}
We present \sys to detect ICS attacks by leveraging unique \emph{execution phases} of SCADA to identify the limited set of benign behaviors to control the physical world in different phases, which differentiates from attacker's activities. To do this, \sys first leverages OPC to build a physical model, which enables it to detect physical anomalies. \sys then uses this model to inform a \emph{physical process-aware} dynamic analysis, whereby SCADA is induced to reveal API calls unique to \emph{process-control}. Through this, \sys detects attack behaviors that violates \emph{process-control} phase.
\sys achieved high accuracy and outperformed existing work.

\section{\xspace \xspace \xspace \xspace ACKNOWLEDGEMENT}
We thank the anonymous reviewers for their helpful and informative feedback. We also thank Marshall Daniels for his guidance. This material was supported in part by Sandia National Laboratories, the Office of Naval Research (ONR) under grants N00014-19-1-2179, N00014-17-1-2895, and N00014-18-1-2662, the Defense Advanced Research Projects Agency (DARPA) under contracts HR00112090031 and HR00112190087. Sandia National Laboratories is a multimission laboratory managed and operated by National Technology \& Engineering Solutions of Sandia, LLC, a wholly owned subsidiary of Honeywell International Inc., for the U.S. Department of Energy National Nuclear Security Administration under contract DE-NA0003525. Any opinions, findings, and conclusions or recommendations expressed in this material are those of the authors and do not necessarily reflect the views of our sponsors. 

\printbibliography[title=References]
\clearpage
\newpage
\appendix
\label{sec:appendix}

\begin{figure}[h]
\centering
	\includegraphics[width=0.38\textwidth]{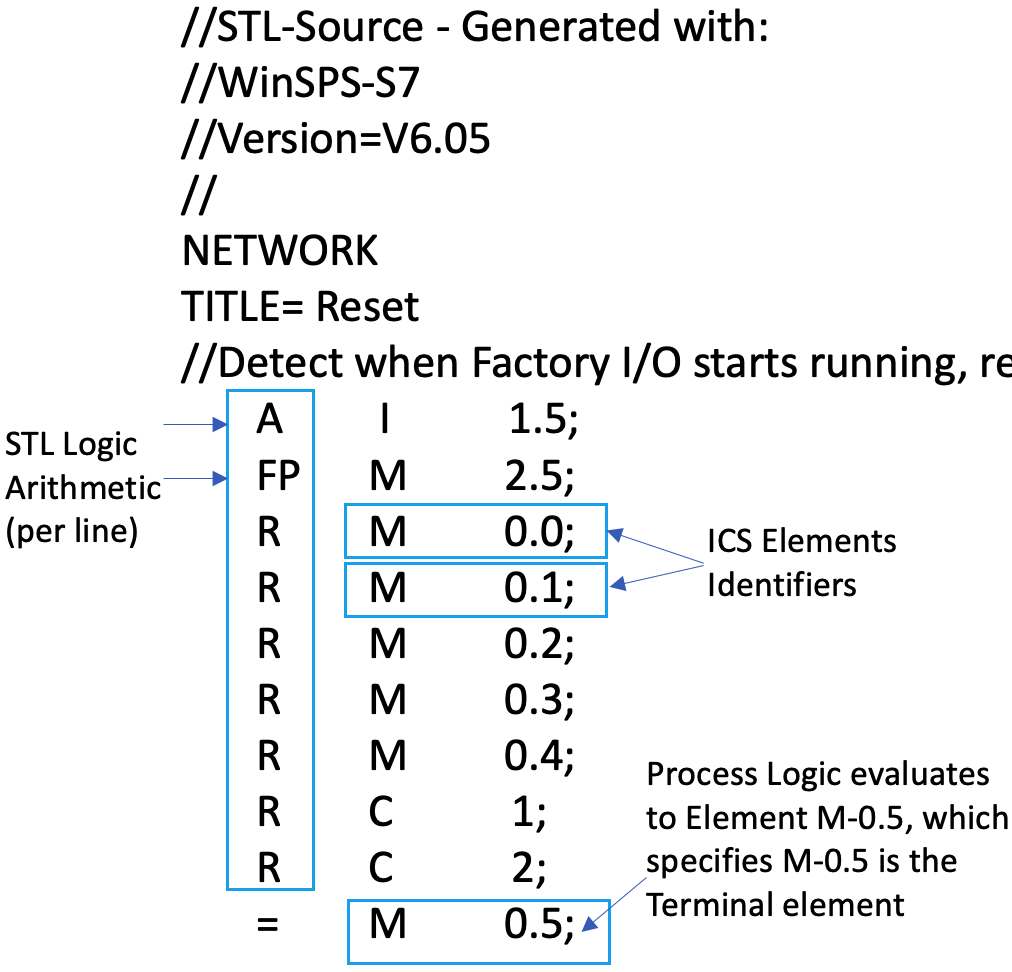}
    \caption{\mr{Our Example Statement List (STL) TEXT FILE Exported from WinSPS-S7 Function Block Diagram. Annotated to show the different parts}}
\label{fig:stl}
\end{figure}

\vspace{3em}

\begin{figure}[h]
\centering
	\includegraphics[width=0.4\textwidth]{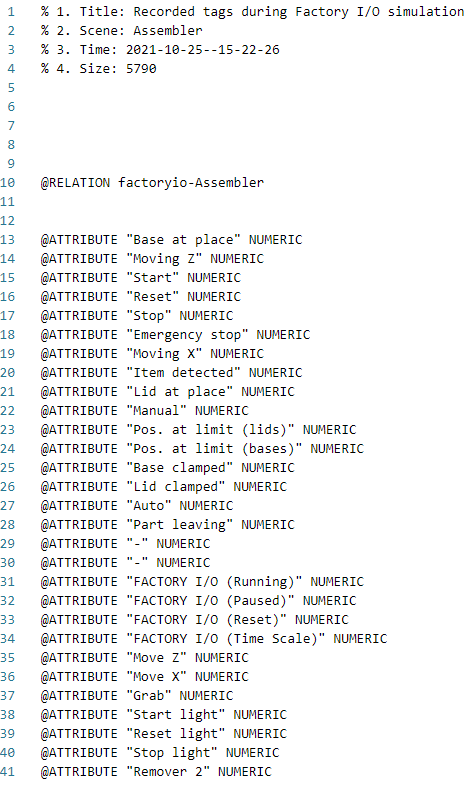}
    \caption{\mr{Our Example ARFF Header showing fields (Element states) we used for specific element in the Sorting Station Scenario}}
\label{fig:arff_header}
\end{figure}
\begin{figure}[h]
\centering
	\includegraphics[width=0.37\textwidth]{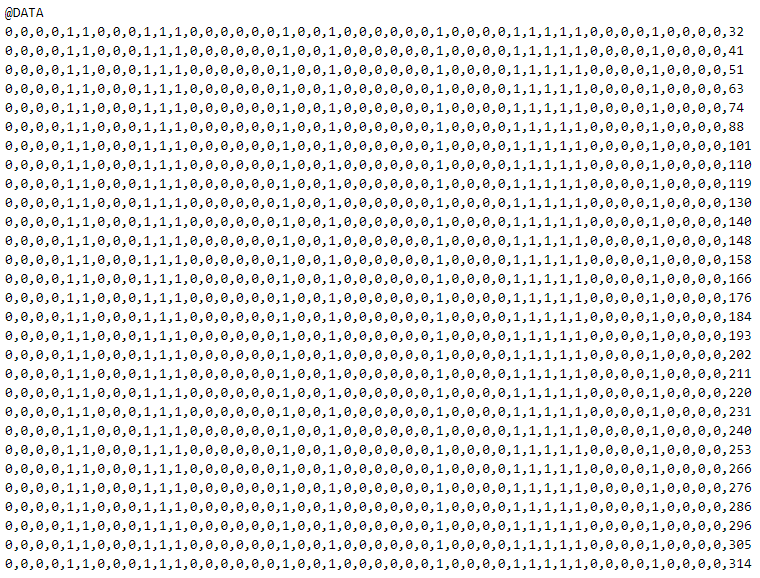}
    \caption{\mr{Sample ARFF-formatted Element State reporting every 100 scan cycles}}
\label{fig:arff_data}
\end{figure}

\begin{figure}[h]
\centering
	\includegraphics[width=0.43\textwidth]{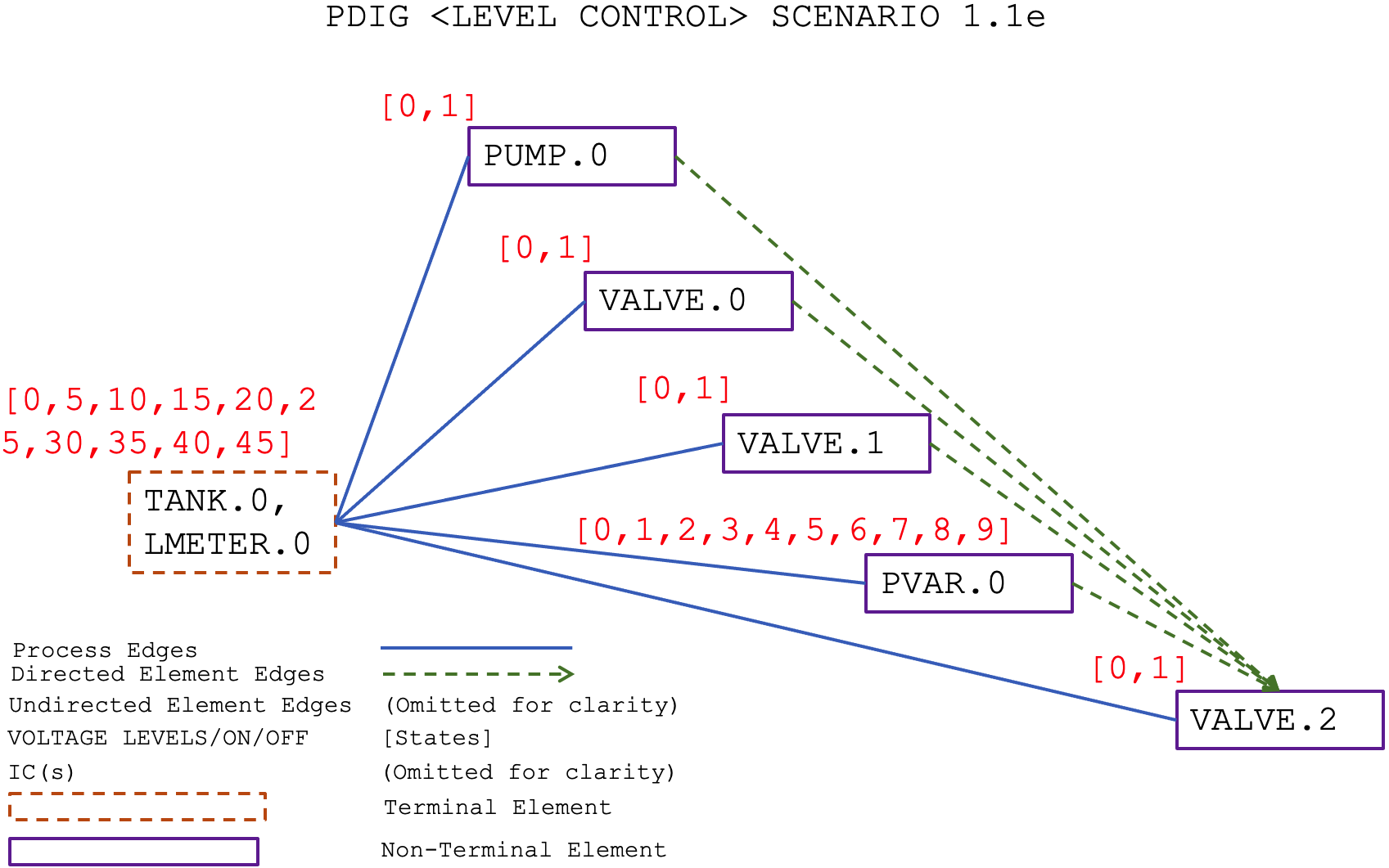}
  \subcaption{\mr{Example PDIG Model for the Level Control Process of the FL Attack Incident}}
\label{fig:pdig}
\end{figure}


\begin{figure}[h]
\centering
	\includegraphics[width=0.49\textwidth]{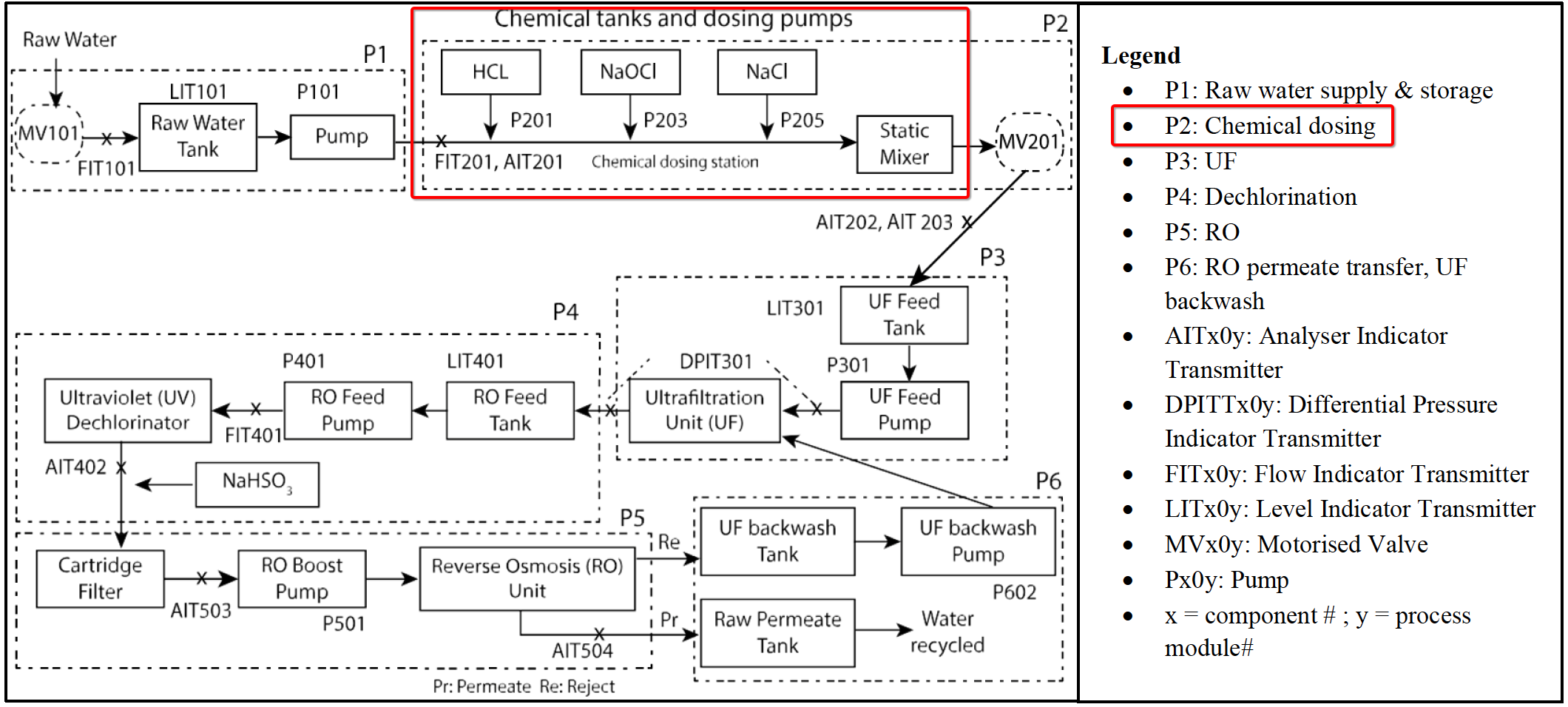}
    \caption{Complete water treatment plant based on~\cite{swat1, swat2}: Showing the chemical dosing operation in reference to the FL water poisoning attack}
\label{fig:complete_water_treatment}
\end{figure}

\end{document}